\documentclass[11pt, preprint]{aastex63}

\usepackage{rotating}
\usepackage{bm}

\newcommand{\ii}{\mathrm{in}}
\newcommand{\oo}{\mathrm{out}}

\newcommand{\kep}{\mathrm{Kep}}

\newcommand{\BBH}{{\rm\scriptscriptstyle BBH}}

\newcommand{\ooz}{{\oo,0}}

\received{2020 September 21}
\revised{2020 November 26}
\accepted{2020 November 28}
\submitjournal{ApJ}

\shorttitle{Pulsar timing analysis of pulsar-binary black hole triples}
\shortauthors{Hayashi \& Suto}
\graphicspath{{./}{figures/}}
\begin{document}

\title{Unveiling the architecture of a pulsar - binary black-hole
  triple system with pulsar arrival time analysis}

\correspondingauthor{Toshinori Hayashi}
\email{toshinori.hayashi@phys.s.u-tokyo.ac.jp}

\author[0000-0003-0288-6901]{Toshinori Hayashi}
\affiliation{Department of Physics, The University of Tokyo,  
Tokyo 113-0033, Japan}

\author[0000-0002-4858-7598]{Yasushi Suto}
\affiliation{Department of Physics, The University of Tokyo,  
Tokyo 113-0033, Japan}
\affiliation{Research Center for the Early Universe, School of Science,
The University of Tokyo, Tokyo 113-0033, Japan}

\begin{abstract}
A large number of binary black holes (BBHs) with longer
  orbital periods are supposed to exist as progenitors of BBH mergers
  recently discovered with gravitational wave (GW) detectors.  In our
  previous papers, we proposed to search for such BBHs in triple
  systems through the radial-velocity modulation of the tertiary
  orbiting star. If the tertiary is a pulsar, high precision and
  cadence observations of its arrival time enable an unambiguous
  characterization of the pulsar -- BBH triples located at several
  kpc, which are inaccessible with the radial velocity of stars. The
  present paper shows that such inner BBHs can be identified through
  the short-term R\o mer delay modulation, on the order of $10$ msec
  for our fiducial case, a triple consisting of $20~M_\odot$ BBH and $1.4~M_\odot$ pulsar with $P_\ii=10$ days and $P_\oo=100$ days. If the relativistic time delays are measured
  as well, one can determine basically all the orbital parameters of
  the triple.  For instance, this method is applicable to inner BBHs
  of down to $\sim 1$ hr orbital periods if the orbital period of the
  tertiary pulsar is around several days.  Inner BBHs with $\lesssim
  1$ hr orbital period emit the GW detectable by future space-based GW
  missions including LISA, DECIGO, and BBO, and very short inner BBHs
  with sub-second orbital period can be even probed by the existing
  ground-based GW detectors. Therefore, our proposed methodology
  provides a complementary technique to search for inner BBHs in
  triples, if exist at all, in the near future.
\end{abstract}

\keywords{celestial mechanics - (stars:) binaries (including multiple): close
  - stars: black holes - (stars:) pulsars: general}

\section{Introduction} \label{sec:intro}

The presence of abundant binary black holes (BBHs) in the
  universe is now firmly established by their gravitational wave (GW)
  signals emitted at the final epoch of their merger event
  \citep[e.g.][]{Abbott2016,Abbott2020}.  The formation channels of
  such BBHs are still controversial, but a variety of scenarios have
  been proposed, including the isolated binary evolution
  \citep[e.g.][]{Belczynski2002}, the dynamical capture in stellar
  systems \citep[e.g.][]{Zwart2000}, and the interaction among
  primordial black holes \citep[e.g.][]{Sasaki2016}.

Regardless of the details of those formation scenarios, the
  merging BBHs indicate that a significantly larger number of
  wide-orbit progenitor BBHs exist, losing their orbital energy before
  coalescence due to the gradual GW emission
  \citep[e.g.][]{Belczynski2016}. Therefore, it is important to
  consider how to detect the population of such wide-separation BBHs
  that are supposed to be hidden even in our Galaxy.

Our previous papers \citep[][hereafter, Papers I and
  II]{Hayashi2020,Hayashi2020_2} proposed a possible methodology to
search for such progenitor BBHs in triple systems by precise radial
velocity (RV) monitoring of a tertiary star orbiting an inner BBH.
In the present paper, we extend the methodology for a triple system
consisting of an inner BBH and a tertiary pulsar on the basis of the
pulsar arrival time analysis.

The formation paths for such triples are highly uncertain, but
  there are several relevant theoretical studies
  \citep[e.g.][]{Toonen2016}.  \citet{Fragione2020}, for instance,
  performed a series of systematic numerical simulations to find the
  demographics of stellar and compact-object triples assembled in
  dense star clusters. They found that binary-binary encounters
  efficiently form triples and that a cluster typically assembles
  hundreds of triples with an inner BBH (of which 70 - 90 \% have a BH
  as tertiary) or an inner star-BH binary.  Thus a triple of an inner
  BBH and a tertiary pulsar may be rare in the scenario.

On the other hand, more than 70 \% of OBA-type stars and 50 \%
of FGK-type stars are known to be in multiples
\citep[][]{Raghavan2010,Sana2012}, and it is likely that yet unknown
dynamical formation channel may account for pulsar - BBH triple
systems.

Indeed, \citet{Ransom2014} discovered an interesting triple
  system with the pulsar timing, which consists of an inner binary of
  a $1.4~M_\odot$ millisecond pulsar (PSR J0337+1715) and $0.2~
  M_\odot$ white dwarf, and a tertiary white dwarf of
  $0.4~M_\odot$. Furthermore, the triple has near-coplanar and circular
  orbits; the mutual inclination $i_\mathrm{mut}$, inner
  eccentricity $e_\ii$, and outer eccentricity $e_\oo$ are
  $1.20\times10^{-2}~\mathrm{deg}$, $6.9178\times10^{-4}$, and
  $3.53561955\times10^{-2}$, respectively.

While the presence of triples consisting of inner BBH and a
  tertiary pulsar is highly uncertain, and their fraction is expected
  to be smaller than star - BBH triples, the tertiary pulsar, if
  exists, is an ideal probe of the architecture of an otherwise unseen
  inner BBH of the triple system. The identification of an inner BBH
  from the precise radial-velocity of the tertiary star is possible
  only when the triple is located very close to us, typically much less
  than $\sim 1$ kpc. In contrast, the pulsar timing method is
  feasible for much more distant triples, even several kpc away. Thus
  the two methodologies may be regarded as complementary.

In this paper, we consider a pulsar orbiting an inner BBH, and
  show the signature and its amplitude of inner BBH hidden in the
  pulsar time of arrival (the R\o mer delay). Furthermore, we point out
  that the other general relativistic effects, the Shapiro and
  Einstein delays, due to the inner BBH, significantly improves the
  understanding of the architecture of the triple system.

The rest of the paper is organized as follows. In section
\ref{sec:method}, we first introduce the overview of the
time-delay effects in a triple. We additionally compare the amplitude
of each effect using the analytic approximation formula for a coplanar
and near-circular triple. Then, we discuss how we can recover the
orbital information of a triple, such as masses and orbital
separation, from detection of each effect in principle.  In section
\ref{sec:example}, we show the time-delay curves of each effect for
three models using the analytic expressions of time-delay
effects.  In section \ref{sec:application}, we put rough constraints on
the currently known double neutron-star binaries from root mean
squares of the observational time residuals. In addition, we discuss
the possibility of our methodology as a complementary technique to the
future direct observation of low-frequency GWs with space-based GW detectors including LISA, DECIGO, and BBO.  Finally, we make a conclusion in section \ref{sec:conclusion}. In the Appendix, we briefly discuss the modulation on
the Keplerian motion induced by the inner-binary perturbation using the analytical formulation of perturbation.

\section{Method}\label{sec:method}

\subsection{Triple configuration}

\begin{figure*}
\begin{center}
\includegraphics[clip,width=10.0cm]{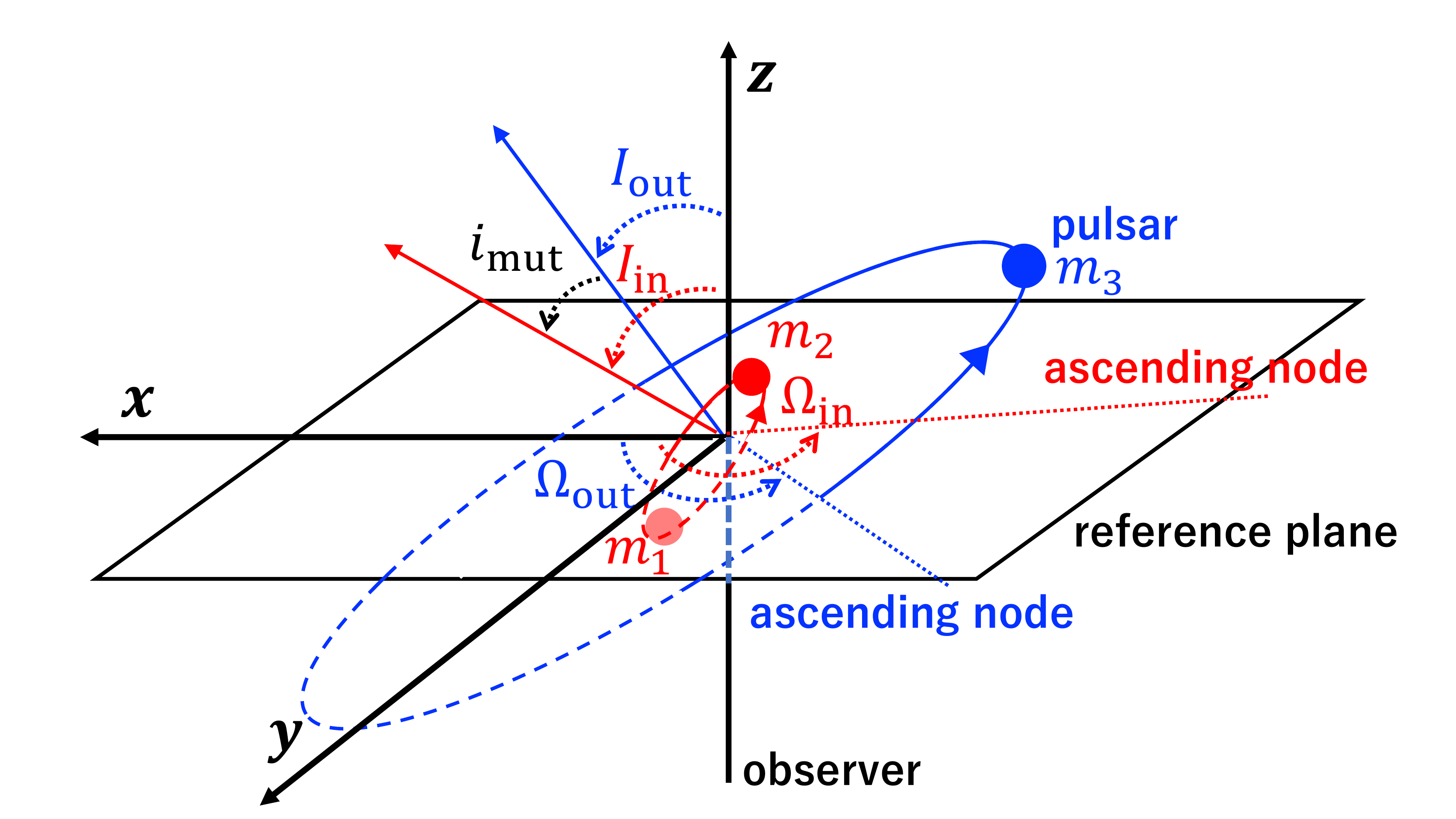}
\end{center}
\caption{ Orbital configuration of a triple system consisting of an
  inner unseen binary and a tertiary millisecond pulsar. The origin of the reference frame is set to be the barycenter of the inner binary, instead of the triple.
  \label{fig:triple}}  
\end{figure*}

Figure \ref{fig:triple} depicts the schematic configuration of a
  hierarchical triple system considered in the present paper.  The
  inner binary comprises two BHs of masses $m_1$ and $m_2$, and is
  orbited by a tertiary pulsar of mass $m_3$.  While we assume the
  inner BBH, the discussion below is equally applicable to other
  compact objects like white dwarfs (WDs) and neutron stars (NSs).

The orbits in the triple system are divided into an inner and outer
orbits in terms of the Jacobi coordinates, and each orbit is
characterized by the semi-major axis $a_j$, the eccentricity $e_j$,
and the argument of pericenter $\omega_j$, where the index $j=\{\ii,
\oo\}$ refers to the inner and outer orbits, respectively. The orientation of each orbit is specified by two angles relative to the
  reference frame; the inclination $I_j$ (the angle between the normal
  vector of the orbit and $z$-axis) and the longitude of ascending
  node $\Omega_j$ measured from $x$-axis.

The mutual inclination between the inner and outer orbits is denoted
by $i_\mathrm{mut}$, and the orbital period and the
corresponding mean motion for each orbit are defined as $P_j$ and
$\nu_j (=2\pi/P_j)$. Throughout the present analysis, we consider a
hierarchical triple system that satisfies $P_\ii \ll P_\oo$.
For definiteness, we assume that the distant observer is
located in the negative $z$-axis.

Following Paper II, we assume a fiducial set of parameters,
$P_\ii=10$ days, $P_\oo=100$ days, and $m_1=m_2=10~M_\odot$, which will
be used in evaluating characteristic amplitudes of the time delays.

\subsection{Pulsar arrival time delays \label{subsec:time-delay}}

The architecture of the triple system is encoded by the motion of
  the tertiary object orbiting the unseen inner binary. If the
  tertiary is a visible star, its radial velocity carries the key
  information (Papers I and II). If the tertiary is a pulsar, its
  arrival time variation has almost identical information of its
  radial velocity, but with much higher precision. Furthermore,
  general relativistic effects provide additional and complementary
  information on the parameters of the triple system.

The arrival time of the pulsar is modulated due to the periodic
  change of the position of the pulsar.  For a pulsar binary system,
  there are three well-known effects including the R\o mer delay
  $\Delta_{\mathrm{R}}(t)$, the Einstein delay
  $\Delta_{\mathrm{E}}(t)$, and the Shapiro delay
  $\Delta_{\mathrm{S}}(t)$. In the case of the hierarchical triple
  system, the main contribution to those three terms comes from the
  approximate binary motion of the pulsar relative to the barycenter
  of the inner binary of mass $m_{12} \equiv m_1+m_2$, whose explicit
  expressions\citep[e.g][]{Backer1986,Edwards2006} are given in this
  subsection.  More importantly, the inner binary produces an
  additional modulation to the tertiary motion of a period
  approximately $P_\ii/2$. Consequently, the R\o mer delay is given by
  $\Delta_{\mathrm{R, Kep}}(t)$ and $\Delta_{\mathrm{R, BBH}}(t)$ that
  can be distinguished from their different frequencies. Indeed the
  inner binary information is imprinted in $\Delta_{\mathrm{R,
      BBH}}(t)$ as we will see below.

\subsubsection{The R\o mer delay due to the Keplerian motion
  of the tertiary pulsar}

The R\o mer delay is caused by the change of the distance of the
  pulsar relative to the observer. To the lowest order, the pulsar
  moves along a Keplerian orbit around the barycenter of the triple
  system.  The corresponding R\o mer delay is written in terms of the
  eccentric anomaly of the outer orbit, $E_\oo=E_\oo(t)$, as
\begin{eqnarray}
\label{eq:Roemer}
\Delta_{\mathrm{R,Kep}}(t)
= x [\sin{\omega_\oo}(\cos{E_\oo}-e_\oo)
  +\sqrt{1-e_\oo^2}\cos{\omega_\oo}\sin{E_\oo}] 
\end{eqnarray}
The eccentric anomaly is expressed implicitly as function of time
  through Kepler's equation:
\begin{eqnarray}
\label{eq:Kepler-equation}
2\pi t = P_\oo (E_\oo -e_\oo \sin E_\oo).
\end{eqnarray}
Note that we can only observe the difference of the time delay, and can choose arbitrarily the zero point of the overall time delays.

Since the semi-major axis of the outer orbit, $a_\oo$, is defined
  with respect to the center of mass of the inner binary, that of the pulsar orbit with respect to the barycenter of the triple is given by $a_{\rm p}= (m_{12}/m_{123})a_\oo$,
  where $m_{12} \equiv m_1+m_2$ and $m_{123} \equiv m_1+m_2+m_3$.
  Thus the amplitude of the R\o mer delay for a distant observer at
  the negative $z$-direction (Fig.\ref{fig:triple}) reduces to
\begin{eqnarray}
\label{eq:Roemer-x}
  x\equiv \frac{m_{12}}{m_{123}}\frac{a_\oo\sin{I_\oo}}{c}
  \approx 570~\mathrm{sec} \left(\frac{m_{12}}{m_{123}}\right)
  \left(\frac{m_{123}}{20~M_\odot}\right)^{1/3}
  \left(\frac{P_\oo}{100~\mathrm{days}}\right)^{2/3}\sin{I_\oo}.
\end{eqnarray}

\subsubsection{The Einstein delay and the Shapiro delay}

In addition to the R\o mer delay, there are two important general
  relativistic effects that carries the information of the triple system;
  the Einstein delay and the Shapiro delay.

The Einstein delay is caused by the difference between the proper time of the pulse emission and the coordinate time of the barycenter of the binary under the gravitational field, and is explicitly expressed as
\begin{eqnarray}
\label{eq:Einstein}
\Delta_\mathrm{E}(t) = \gamma_\mathrm{E} \sin{E_\oo},
\end{eqnarray}
where the amplitude is given by
\begin{eqnarray}
\label{eq:Einstein-amplitude}
\gamma_\mathrm{E} &\equiv& \left(\frac{\mathcal{G}}{c^3}\right)^{2/3}
\left(\frac{P_\oo}{2\pi}\right)^{1/3}e_\oo
\frac{m_{12}(m_3+2m_{12})}{m_{123}^{4/3}} \cr
&\approx & 2.4~\mathrm{msec}
\left(\frac{P_\oo}{100~\mathrm{days}}\right)^{1/3}
\left(\frac{m_{12}}{20~M_\odot}\right)
\left(\frac{m_{123}}{20~M_\odot}\right)^{-1/3}
\left(1+\frac{m_{12}}{m_{123}}\right)\left(\frac{e_\oo}{0.01}\right)
\end{eqnarray}
with $\mathcal{G}$ being the gravitational constant.

The Shapiro delay is the time delay during the passage of the photon
due to the gravitational space-time curvature of the companion, and is
given by
\begin{eqnarray}
\label{eq:Shapiro}
\Delta_\mathrm{S}(t) &=& -2r \ln\left[1-e_\oo\cos{E_\oo}\right. \cr
  &&\qquad\qquad\qquad
  \left. - s \left(\sin{\omega_\oo}(\cos{E_\oo}-e_\oo)
  +\sqrt{1-e_\oo^2}\cos{\omega_\oo}\sin{E_\oo}\right)\right],
\end{eqnarray}
where $r$ and $s$ are the major observables and usually referred to as
``range'' and ``shape'' parameters\citep[e.g][]{Backer1986,Edwards2006}:
\begin{eqnarray}
  \label{eq:range-parameter}
  r &\equiv& \frac{\mathcal{G}m_{12}}{c^3}
  \approx 98~ \mathrm{\mu sec} \left(\frac{m_{12}}{20~M_\odot}\right), \\
  \label{eq:shape-parameter}
  s &\equiv& \sin{I_\oo} .
\end{eqnarray}

As equations (\ref{eq:Einstein}) and (\ref{eq:Shapiro}) indicate,
  $\Delta_\mathrm{S} \ll \Delta_\mathrm{E}$ except for for a nearly
  circular and edge-on system ($e_\oo\approx 0$ and $s \approx 1$). A
  notable example is a binary pulsar system PSR J1614-2230 with $e_\oo
  = 1.30\times10^{-6}$ and $I_\oo = 89.17$ deg, which reveals the
  mass of the pulsar is as massive as $1.97~M_\odot$ via the analysis
  of the Shapiro delay \citep[][]{Demorest2010}.
  
In principle, the Shapiro delays for both inner and outer orbits
    may be separately detected, especially
    for nearly edge-on coplanar systems ($I_\oo\approx I_\ii \approx \pi/2$).
    In this case, the Shapiro delays alone
    directly reveal the existence of the inner binary and their
    individual masses $m_1$ and $m_2$. In what follows, however,
    we conservatively consider only the Shapiro delay for the outer orbit,
    assuming the inner binary as a single object of mass $m_{12}$.

\subsubsection{The R\o mer delay due to the inner binary motion}

The orbital motion of the inner BBH perturbs the Keplerian motion
  of the tertiary. Papers I and II proposed to detect the induced
  radial velocity modulation of the tertiary star to search for a
  possible inner BBH in the triple systems.  In the case of a coplanar
  and near-circular hierarchical triple, the corresponding radial
  velocity variation up to the quadrupole order of the BBH potential
  is approximately given by
\citep[][see also paper I and II]{Morais2008}
\begin{eqnarray}
  \label{eq:zdot-BBH}
  \dot{z}_\mathrm{BBH}(t) = \frac{15}{16} K_\BBH
  \sin{I_\oo}\cos(\nu_{-3}t+\theta_{0,-3})
  + \frac{3}{16} K_\BBH \sin{I_\oo}\cos(\nu_{-1}t+\theta_{0,-1}),
\end{eqnarray}
where the characteristic velocity amplitude is
\begin{eqnarray}
  K_\BBH \equiv \frac{m_1m_2}{m_{12}^2} \sqrt{\frac{m_{123}}{m_{12}}}
  \left(\frac{a_\ii}{a_\oo}\right)^{7/2}\times
  \left[\frac{m_{12}}{m_{123}} a_\oo \nu_\oo\right]
\end{eqnarray}
with the term inside the square bracket corresponding to the velocity of the circular Keplerian motion, and the frequencies of the two modes are
\begin{eqnarray}
  \nu_{-3} &\equiv& 2\nu_\ii-3\nu_\oo \\
  \nu_{-1} &\equiv& 2\nu_\ii-\nu_\oo,
\end{eqnarray}
with $\theta_{0,-3}$ and $\theta_{0,-1}$ being the initial phases.
For the hierarchical triples ($\nu_\ii \gg \nu_\oo$), $\nu_{-3}
  \approx \nu_{-1} \approx 2\nu_\ii =4\pi/P_\ii$. Thus the
  high-cadence radial velocity observation is required to break the
  degeneracy of the two modes.

Equation (\ref{eq:zdot-BBH}) is directly translated to
  the short-term R\o mer delay:
\begin{eqnarray}
\label{eq:Roemer-BBH}
\Delta_\mathrm{R,BBH}(t) \equiv \frac{z_\BBH(t)}{c}
&=& \frac{15}{16}\frac{K_\BBH P_\ii}{4\pi c}
\left(\frac{2\nu_\ii}{\nu_{-3}}\right) \sin{I_\oo}\sin(\nu_{-3}t+\theta_{0,-3}) \cr
&+&
\frac{3}{16}\frac{K_\BBH P_\ii}{4\pi c} \left(\frac{2\nu_\ii}{\nu_{-1}}\right)\sin{I_\oo}\sin(\nu_{-1}t+\theta_{0,-1}).
\end{eqnarray}
The characteristic amplitude of equation (\ref{eq:Roemer-BBH})
is
\begin{eqnarray}
\label{eq:Roemer-BBH-amplitude}
\frac{K_\BBH P_\ii}{4\pi c} \sin{I_\oo} &=& 
\frac{1}{2}\frac{m_1m_2}{m_{12}^2}
\left(\frac{m_{12}}{m_{123}}\right)^{2/3} \left(\frac{P_\ii}{P_\oo}\right)^{7/3}x \cr
&\approx& 23~\mathrm{msec} \left(\frac{K_\BBH}{100~\mathrm{m/s}}\right)
\left(\frac{P_\ii}{10~\mathrm{days}}\right) \sin{I_\oo},
\end{eqnarray}
implying that the inner BBH motion is much larger than the general
relativistic terms in general. Furthermore, the high-cadence
monitoring of the pulsar timing can break the degeneracy between the
$\nu_{-3}$ and $\nu_{-1}$ modes more easily than in the radial
velocity measurements of main-sequence stars.

\begin{figure*}
\begin{center}
\includegraphics[clip,width=8cm]{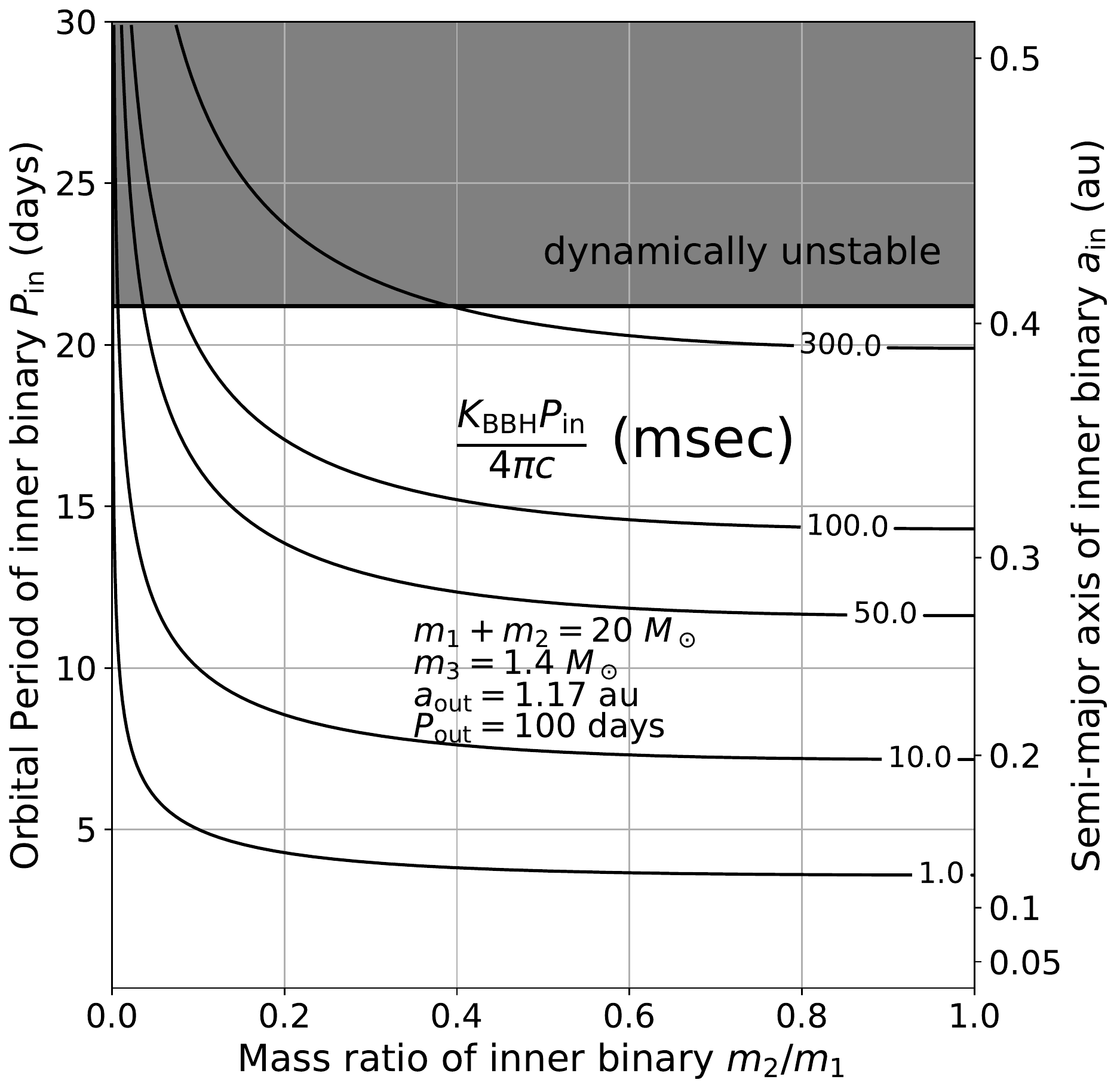}
\end{center}
\caption{Characteristic amplitude of the R\o mer delay induced by
    the inner BBH as a function of the inner-binary orbital period
    ($P_\ii$) and mass ratio ($m_2/m_1$). We assume a coplanar and
    circular triple system with the following parameters; the total
    mass of the inner BBH $m_1+m_2=20~M_\odot$, the tertiary pulsar
    mass $m_3=1.4~M_\odot$, the outer orbital period $P_\oo=100$ days
    corresponding to the outer semi-major axis of $a_\oo=1.17$ au).
    The upper shaded region is dynamically unstable from the criterion
    by \citet{Mardling1999}.
  \label{fig:contour1}}  
\end{figure*}

\begin{figure*}
\begin{center}
\includegraphics[clip,width=8.5cm]{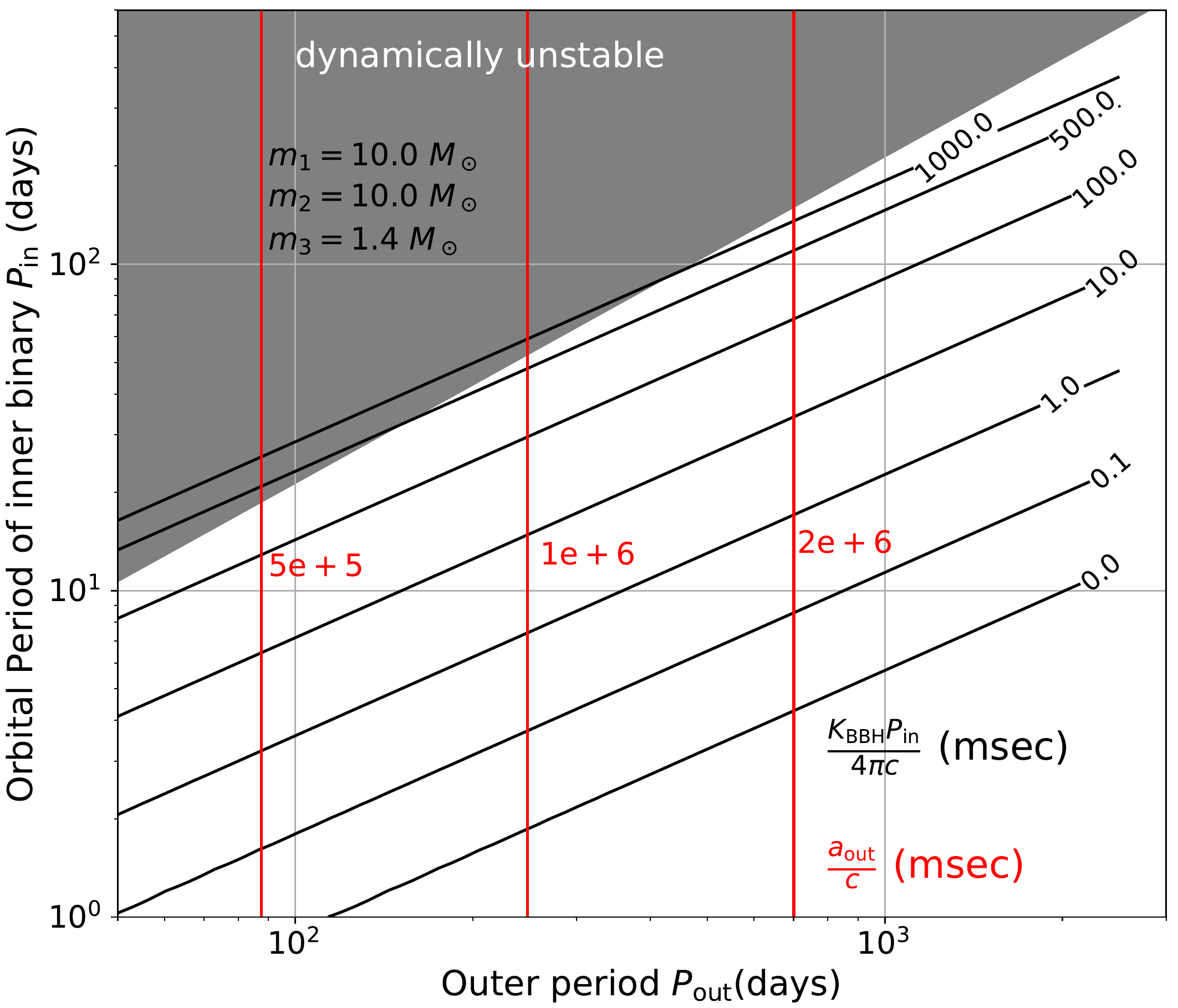}
\end{center}
\caption{Characteristic amplitude of the R\o mer delay induced by the
  equal-mass inner-BBH as a function of the inner and outer orbital
  periods. We assume the same parameters for the triple system as in
  Figure \ref{fig:contour1} except that $m_1=m_2=10~M_\odot$ and
  $P_\oo$ is a variable. For reference, the R\o mer delay amplitude
  due to the tertiary Keplerian motion is plotted in red lines.
    \label{fig:contour2}}  
\end{figure*}

Figures \ref{fig:contour1} and \ref{fig:contour2} show the contour plots of
  the characteristic amplitude of the R\o mer delay induced by the
  inner BBH, equation (\ref{eq:Roemer-BBH-amplitude}), for the case of
  a coplanar and near-circular triple with our fiducial set of
  parameters.  Figure \ref{fig:contour1} is plotted on the $m_2/m_1$ and
  $P_\ii$ plane for $P_\oo=100$ days, while Figure \ref{fig:contour2} is
  plotted on the $P_\oo$ and $P_\ii$ plane for $m_1=m_2=10~M_\odot$.
  Since typical amplitudes of the pulsar timing noise is less than
  $O(100)~\mu$sec (see, e.g., Table \ref{tab:DNS}), the R\o mer delay
  induced by the inner BBH for our fiducial triple may be
  easily detected
  as long as precise and high-cadence pulsar-timing data
  are available.

\begin{figure*}
\begin{center}
\includegraphics[clip,width=7cm]{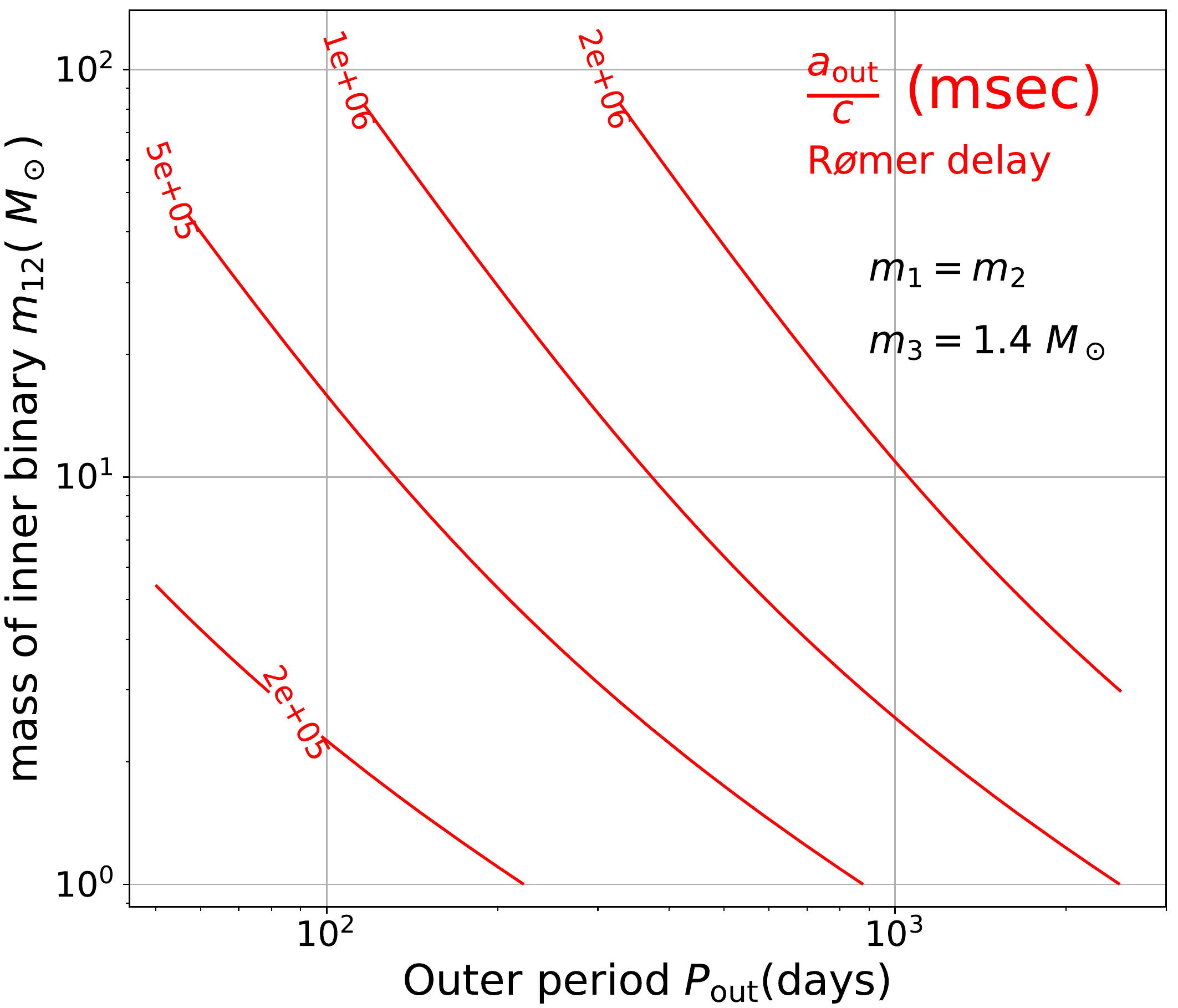}
\hspace{15pt}
\includegraphics[clip,width=7cm]{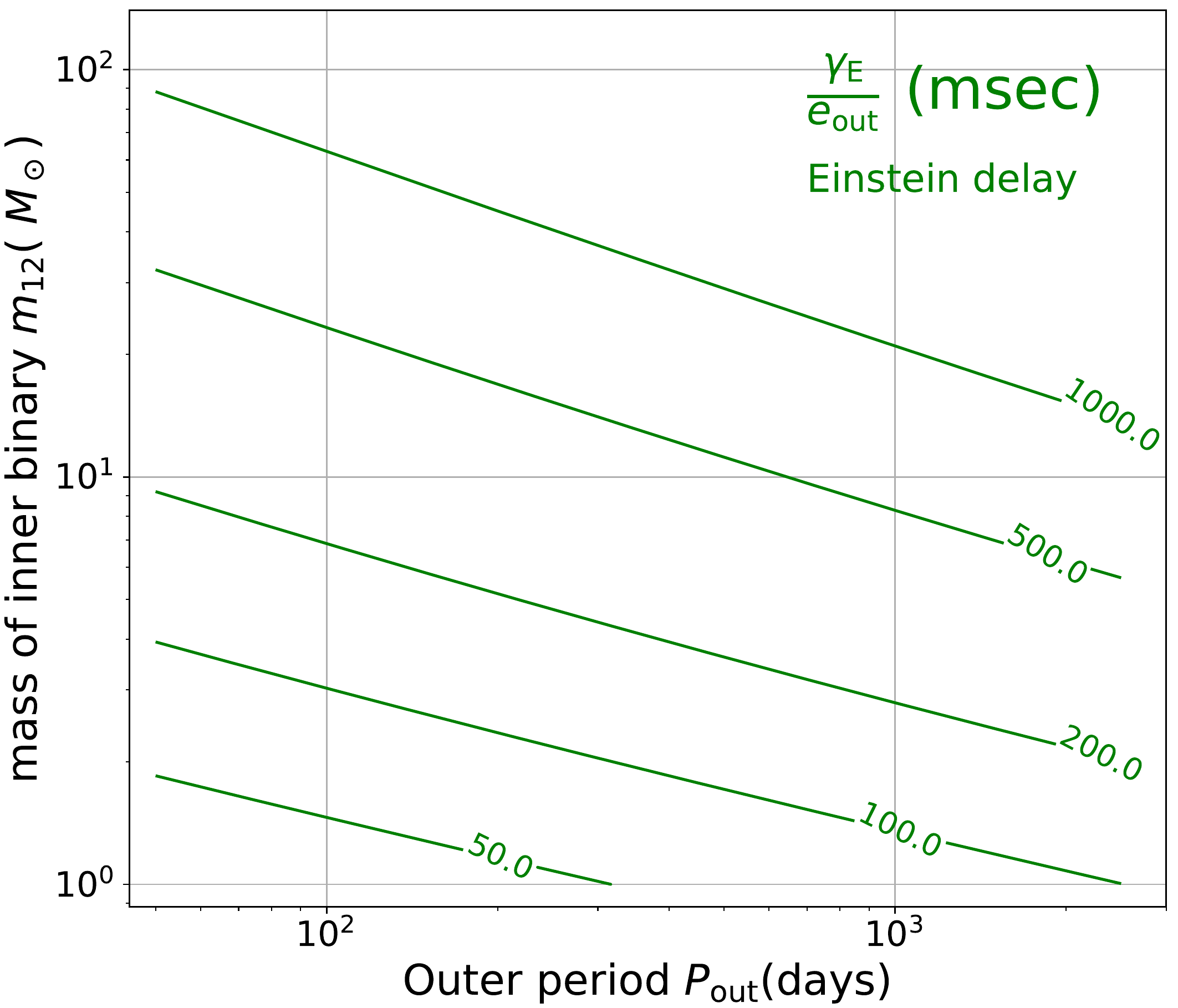}
\hspace{15pt}
\includegraphics[clip,width=7cm]{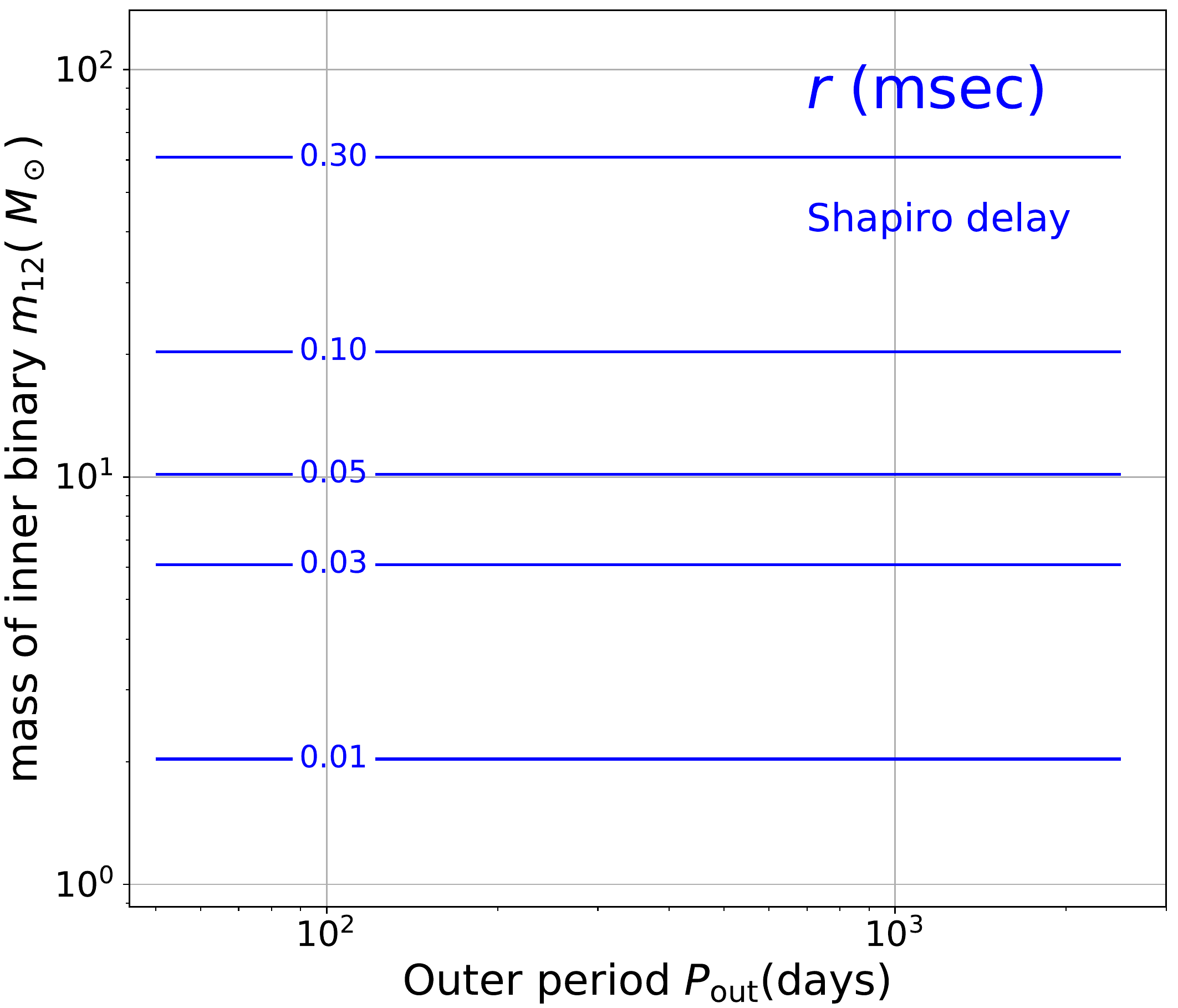}
\hspace{15pt}
\includegraphics[clip,width=7cm]{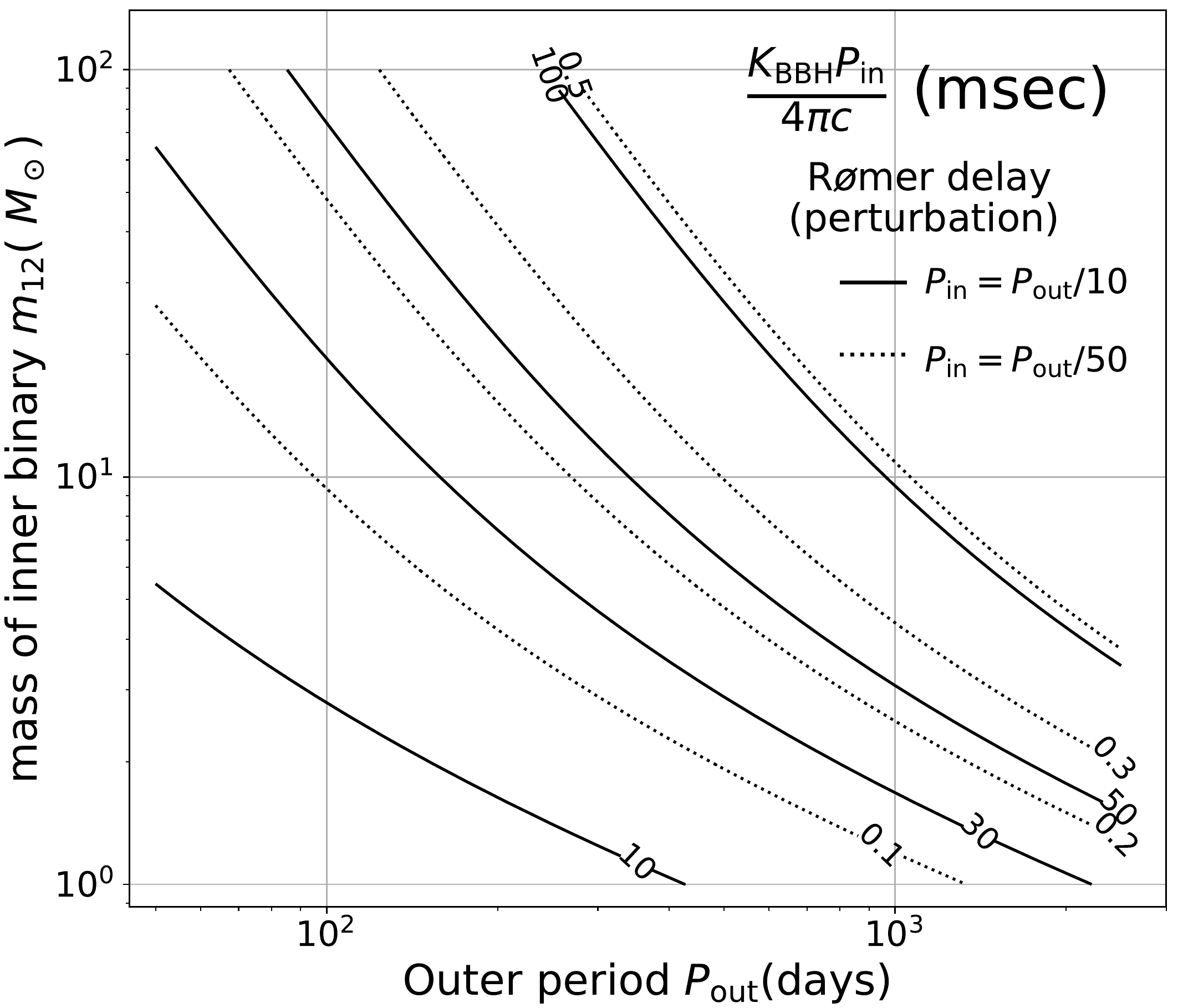}
\end{center}
\caption{Characteristic amplitudes of each time-delay effect as a
  function of the outer orbital period $P_\oo$ and total mass of the
  inner BBH $m_{12}$. The upper-left, upper-right, lower-left, and
  lower-right panels plot the R\o mer delay of the Keplerian motion,
  the Einstein delay, the Shapiro delay, and the R\o mer delay by the
  inner binary perturbation, respectively. The solid and dotted black
  lines in the lower-right panel show the R\o mer time delay induced
  by the inner binary with $P_\ii$ being $P_\oo/10$ and $P_\oo/50$,
  respectively. Note that the amplitude of the Shapiro delay
    is sensitive to the inclination $I_\oo$, or equivalently to the
    shape parameter $s=\sin I_\oo$ (see Figure \ref{fig:time-delay},
    for example). Thus, the range parameter $r$ is just a very crude
    estimate of the amplitude.} \label{fig:contour3}
\end{figure*}

Figure \ref{fig:contour3} compares the characteristic amplitudes of
the four time delays as a function of the mass $m_{12}$ and the outer
orbital period $P_\oo$ for an equal-mass inner binary with the
tertiary mass of $m_3=1.4~M_\odot$. The upper-left, upper-right,
lower-left, and lower-right panels show the R\o mer delay of the outer
Keplerian motion, the Einstein delay, the Shapiro delay, and the R\o
mer delay induced by the inner binary perturbation, respectively.  The
solid and dotted black lines in the lower-right panel show the contour
curves for the cases that $P_\ii=P_\oo/10$, $P_\oo/50$,
respectively. Note that the amplitude of the Shapiro delay is
  very sensitive to the shape parameter $s=\sin I_\oo$ as indicated by
  equation (\ref{eq:Shapiro}). Therefore, the amplitude of the range
  parameter $r$ plotted in Figure \ref{fig:contour3} should be
  regarded as a very rough estimate of the expected Shapiro delay.

\subsection{Constraints on parameters from individual time delay data
  \label{subsec:parameter1}}

\begin{figure*}
\begin{center}
\includegraphics[clip,width=16.5cm]{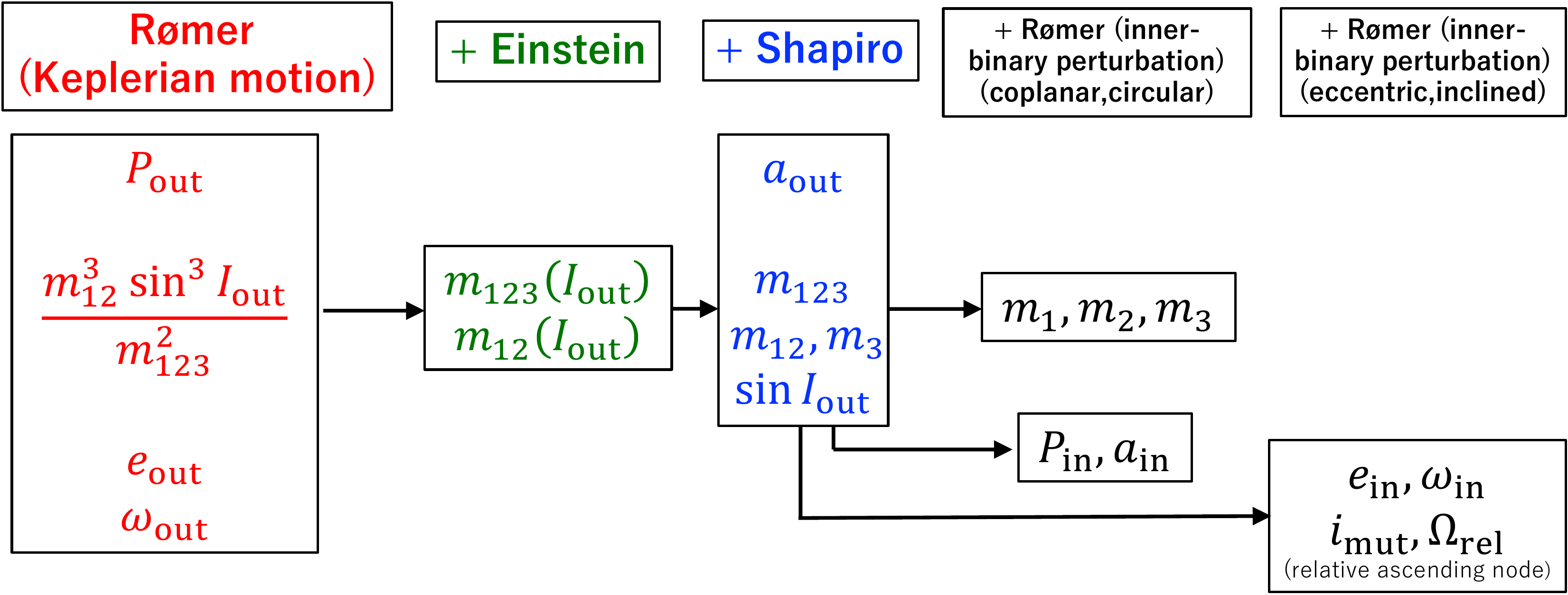}
\end{center}
\caption{Summary of orbital parameters that are estimated from
  the pulsar arrival timing analysis.
  \label{fig:parameter-estimate}}  
\end{figure*}

There are 13 parameters (three masses and ten orbital parameters except for the initial positions of the bodies) in total that specify the triple system,
  and the four time delays discussed in the above subsection put
  different constraints on those parameters. We show those constraints
  separately for each time delay measurement in this subsection, and
  consider how their combination determines the properties of the inner
  BBH in the next subsection. The procedure to estimate orbital
  parameters for the triple system with a tertiary pulsar is summarized
  in Figure \ref{fig:parameter-estimate} 

Consider first the R\o mer time delay of the Keplerian motion.
Strictly speaking, all the orbital parameters in equation
  (\ref{eq:Roemer}) are not constant except for the two-body system.
  For the hierarchical triple system that we consider here, those
  parameters typically vary over the Kozai-Lidov timescale:
\begin{eqnarray}
\label{eq:ta-KL}
\tau_{\rm KL} \equiv
\frac{m_{12}}{m_3} \frac{P_\oo}{P_\ii} P_\oo .
\end{eqnarray}
Throughout the present paper, we assume that the duration of the
  pulsar arrival time data is much less than $\tau_{\rm KL}$, and that
  the orbital parameters are approximately constant. Even if it is not
  the case, however, the methodology that we propose here remains the
  same, but it requires numerical integration of the three-body
  dynamics so as to properly account for the time-dependence of the
  orbital parameters.

Under the approximation, fitting equation (\ref{eq:Roemer}) to
  the series of the pulsar arrival time data determines the values of
  $e_\oo$, $\omega_\oo$, and $P_\oo$, in addition to the overall
  amplitude $x=m_{12}a_\oo\sin{I_\oo}/(cm_{123})$. Combining Kepler's
  third law $a_\oo^3/P_\oo^2=(\mathcal{G}m_{123}/4\pi^2)$, one obtains
  the following relation:
\begin{eqnarray}
\label{eq:binarymassfunc}
  \frac{m_{12}^3\sin^3{I_\oo}}{(m_{12}+m_{3})^2}=(cx)^3\mathcal{G}^{-1}
  \left(\frac{2\pi}{P_\oo}\right)^2.
\end{eqnarray}
Indeed, this is the binary mass function expressed in the
  observables from the R\o mer delay measurement.

Second, the Einstein delay, equation (\ref{eq:Einstein}), yields
  $e_\oo$ and $P_\oo$ through equation (\ref{eq:Kepler-equation}), and
  $\gamma_\mathrm{E}$. From the three parameters, one obtains
\begin{eqnarray}
\label{eq:mass_func2}
\frac{m_{12}(m_{123}+m_{12})}{m_{123}^{4/3}}
= \frac{\gamma_\mathrm{E}}{e_\oo}
\left(\frac{2\pi}{P_\oo}\right)^{1/3}
\left(\frac{\mathcal{G}}{c^3}\right)^{-2/3}.
\end{eqnarray}
Equation (\ref{eq:mass_func2}) is another useful constraint
  on $m_{123}$ and $m_{12}$ that is independent of $I_\oo$ unlike
  equation (\ref{eq:binarymassfunc}) from the R\o mer delay.

Third, the Shapiro delay, equation (\ref{eq:Shapiro}), is
  particularly useful to determine $I_\oo (= \sin^{-1} s)$, in
  addition to $e_\oo$, $P_\oo$, and $r$. Also the range parameter $r$,
  equation (\ref{eq:range-parameter}), is directly related to the
  total mass of the inner BBH binary:
\begin{eqnarray}
m_{12} = r\left(\frac{\mathcal{G}}{c^3}\right)^{-1}.
\end{eqnarray}
Thus, the detection of the Shapiro delay plays a crucial and
complementary role in breaking the degeneracy of the parameter
estimation, especially for systems with $\sin I_\oo \approx 1$.
We also emphasize that the individual detection of the Shapiro
  delays for both components of the inner orbit would clearly break the
  degeneracy of the triple architecture of the system. This is likely to be the case if the triple is a nearly coplanar and
  edge-on system, and one can separately estimate the mass of the
  inner binary $m_1$ and $m_2$ from the Shapiro delays alone.

So far the above observables are mainly for the outer orbital
  parameters, and the R\o mer delay by inner binary motion is the key
  observable to unveil the properties of the inner BBH.  In order to
  show an specific example, we consider a coplanar near-circular
  triple expressed in equation (\ref{eq:Roemer-BBH}).  Fitting
  equation (\ref{eq:Roemer-BBH}) to the pulsar arrival timing data
  yields
\begin{eqnarray}
P_\ii = \frac{8\pi}{3\nu_{-1}-\nu_{-3}}
\end{eqnarray}
and
\begin{eqnarray}
  \frac{m_1m_2}{m_{12}^2}\left(\frac{a_\ii}{a_\oo}\right)^{7/2}
  \sqrt{\frac{m_{123}}{m_{12}}}
  = \frac{K_\BBH}{cx}\left(\frac{P_\oo}{2\pi}\right).
\end{eqnarray}

\subsection{Inner-binary parameters from joint analysis
  of time delays \label{subsec:parameter2}}

In this section, we show how the orbital parameters of an unseen
inner BBH can be recovered from the joint analysis of the
pulsar arrival time. 

As discussed in the previous subsection, the dominant
  contribution comes from the R\o mer delay from the Keplerian motion
  of the outer orbit, which derives $e_\oo$, $P_\oo$, $\omega_\oo$,
  and $x$. If the Einstein delay is detected as well, the total mass
  of the system $m_{123}$ and inner binary mass $m_{12}$ are written
  in terms of the observables and $\sin I_\oo$ from combining
  equations (\ref{eq:binarymassfunc}) and (\ref{eq:mass_func2}):
\begin{eqnarray}
  \label{eq:m123}
  m_{123} = \left(\frac{\mathcal{G}}{c^3}\right)^{-1}
  \frac{1}{x^{3}}
  \left(\frac{P_\oo}{2\pi}\right)
  \left[\frac{\gamma_\mathrm{E}}{e_\oo}
    -\frac{x^2}{\sin^{2}{I_\oo}}
    \left(\frac{P_\oo}{2\pi}\right)^{-1}\right]^3
\end{eqnarray}
and
\begin{eqnarray}
  \label{eq:m12}
m_{12} = \left(\frac{\mathcal{G}}{c^3}\right)^{-1} \frac{1}{x\sin{I_\oo}}
\left[\frac{\gamma_\mathrm{E}}{e_\oo}-\frac{x^2}{\sin^{2}{I_\oo}}
  \left(\frac{P_\oo}{2\pi}\right)^{-1}\right]^2.
\end{eqnarray}

In addition, if the Shapiro delay is detected, $\sin I_\oo$ and
  $m_{12}$ are derived directly from the range and shape parameters
  $r$ and $s$, respectively. Thus equation (\ref{eq:m12}) provides a
  consistency relation among observables:
\begin{eqnarray}
  \label{eq:consistency}
r = \frac{1}{sx}
\left[\frac{\gamma_\mathrm{E}}{e_\oo}-\frac{x^2}{s^2}
  \left(\frac{P_\oo}{2\pi}\right)^{-1}\right]^2.
\end{eqnarray}

Using equation (\ref{eq:consistency}), equation (\ref{eq:m123})
  is rewritten in terms of the observables alone:
\begin{eqnarray}
\label{eq:m123-2}
  m_{123} =   \left(\frac{\mathcal{G}}{c^3}\right)^{-1}
\left(\frac{sr}{x}\right)^{3/2} \left(\frac{P_\oo}{2\pi}\right) .
\end{eqnarray}
Therefore, the mass of the tertiary can be estimated as
\begin{eqnarray}
\label{eq:m3}
m_3 = m_{123}-m_{12} =
  \left(\frac{\mathcal{G}}{c^3}\right)^{-1}
  \left[\left(\frac{sr}{x}\right)^{3/2}
    \left(\frac{P_\oo}{2\pi}\right) -r\right].
\end{eqnarray}
Since the mass of a neutron star is $(1-2)~M_\sun$, equation
  (\ref{eq:m3}) may be also used as a consistency check of the
  analysis. Similarly
the semi-major axis of the outer orbit can be written as
\begin{eqnarray}
a_\oo = c\left(\frac{P_\oo}{2\pi}\right)\left(\frac{sr}{x}\right)^{1/2}. 
\end{eqnarray}

Finally, the R\o mer delay by the inner binary perturbation, if
  observed at all, elucidates the inner orbital parameters from the
  inner orbital period $P_\ii$, and the velocity variation amplitude
  $K_\BBH$; see equation (\ref{eq:Roemer-BBH}).

Specifically, each mass of inner binary components $m_{1,2}$ and
  the inner orbital semi-major axis $a_\ii$ are written as follows:
\begin{eqnarray}
a_\ii = c r^{1/3} \left(\frac{P_\ii}{2\pi}\right)^{2/3}
\end{eqnarray}
and 
\begin{eqnarray}
  m_{1,2} = \left(\frac{\mathcal{G}}{c^3}\right)^{-1}
  \left(\frac{r}{2}\right)
  \left[1 \pm \sqrt{1 - \frac{4K_\BBH}{c}
      \left(\frac{2\pi r}{P_\ii}\right)^{7/3}
      \left(\frac{P_\oo}{2\pi}\right)^4 (rx)^{-2}}\right].
\end{eqnarray}
Note that the above equations are written in terms of the
  observables from the R\o mer delays of the Keplerian motion and
  inner binary perturbation, and the Shapiro delay, but without
the Einstein delay measurement.

\section{Effects of the eccentricity
    and inclination of the inner binary
    on the pulsar arrival time\label{sec:example}}

While the analytic discussion presented in the previous section
  assumes a coplanar near-circular triple, the procedure of the triple
  parameter extraction is the same except that the evolution of the
  triple system needs to be computed numerically in general.  We
  demonstrate the eccentricity and inclination effects on the pulsar
  arrival time separately in this section using the approximate
  analytic formulae by \citet{Morais2011}.

For that purpose, we consider three models listed in Table
  \ref{tab:models}; a coplanar circular triple (model CC), a
coplanar eccentric triple (model CE), and an inclined circular triple
(model IC).

\begin{figure*}
\begin{center}
\includegraphics[clip,width=8.0cm]{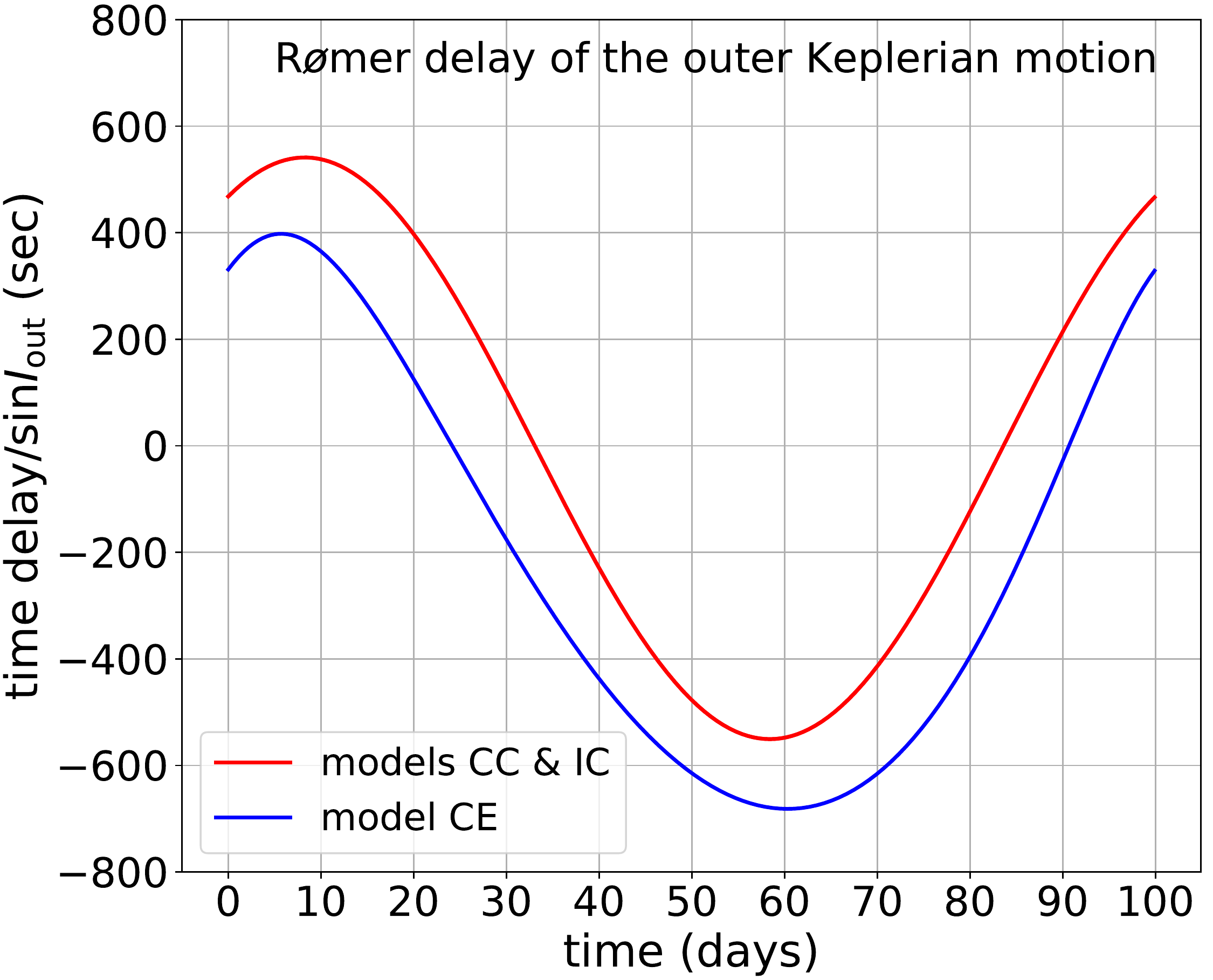}
\hspace{15pt}
\includegraphics[clip,width=8.0cm]{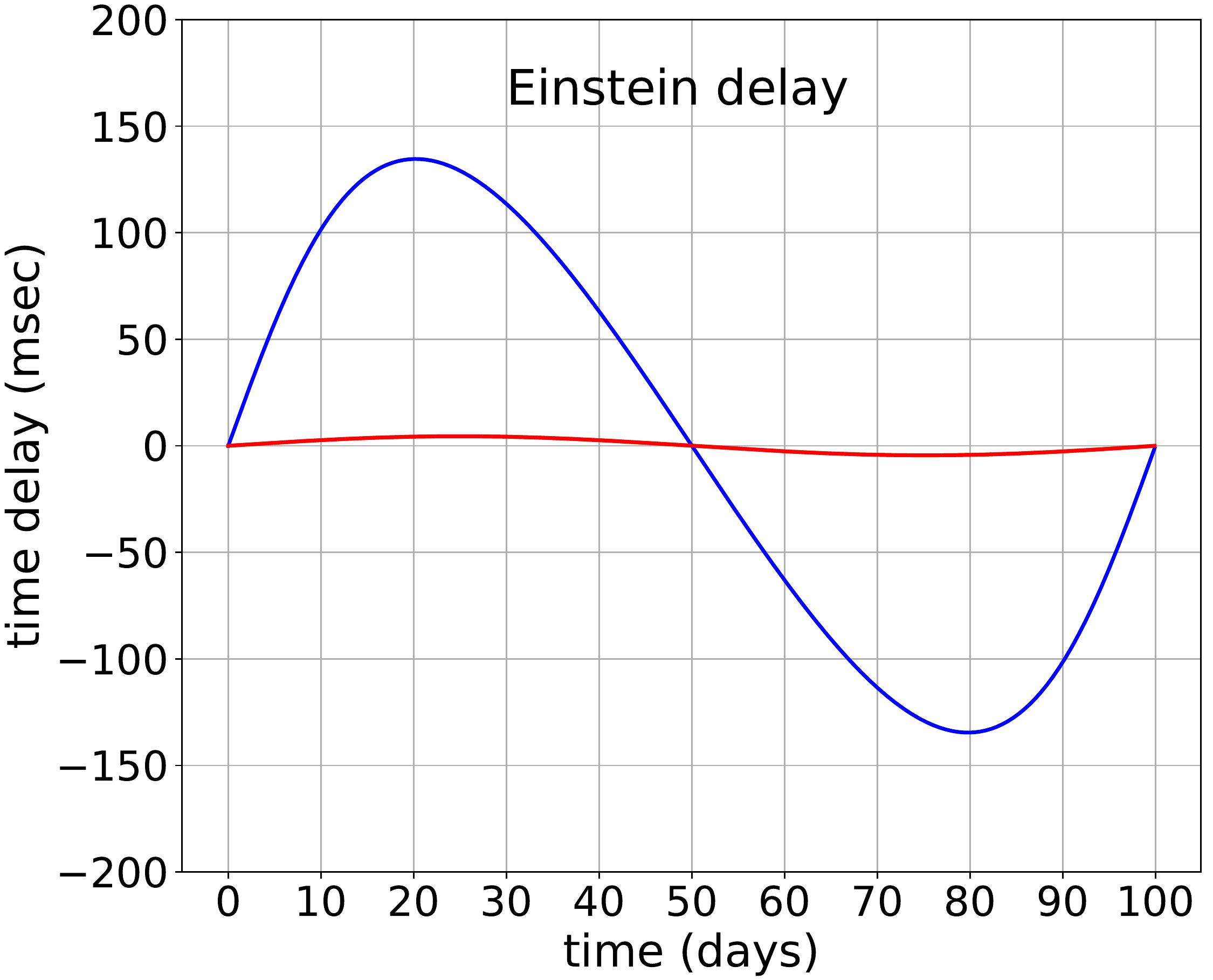}
\hspace{15pt}
\includegraphics[clip,width=8.0cm]{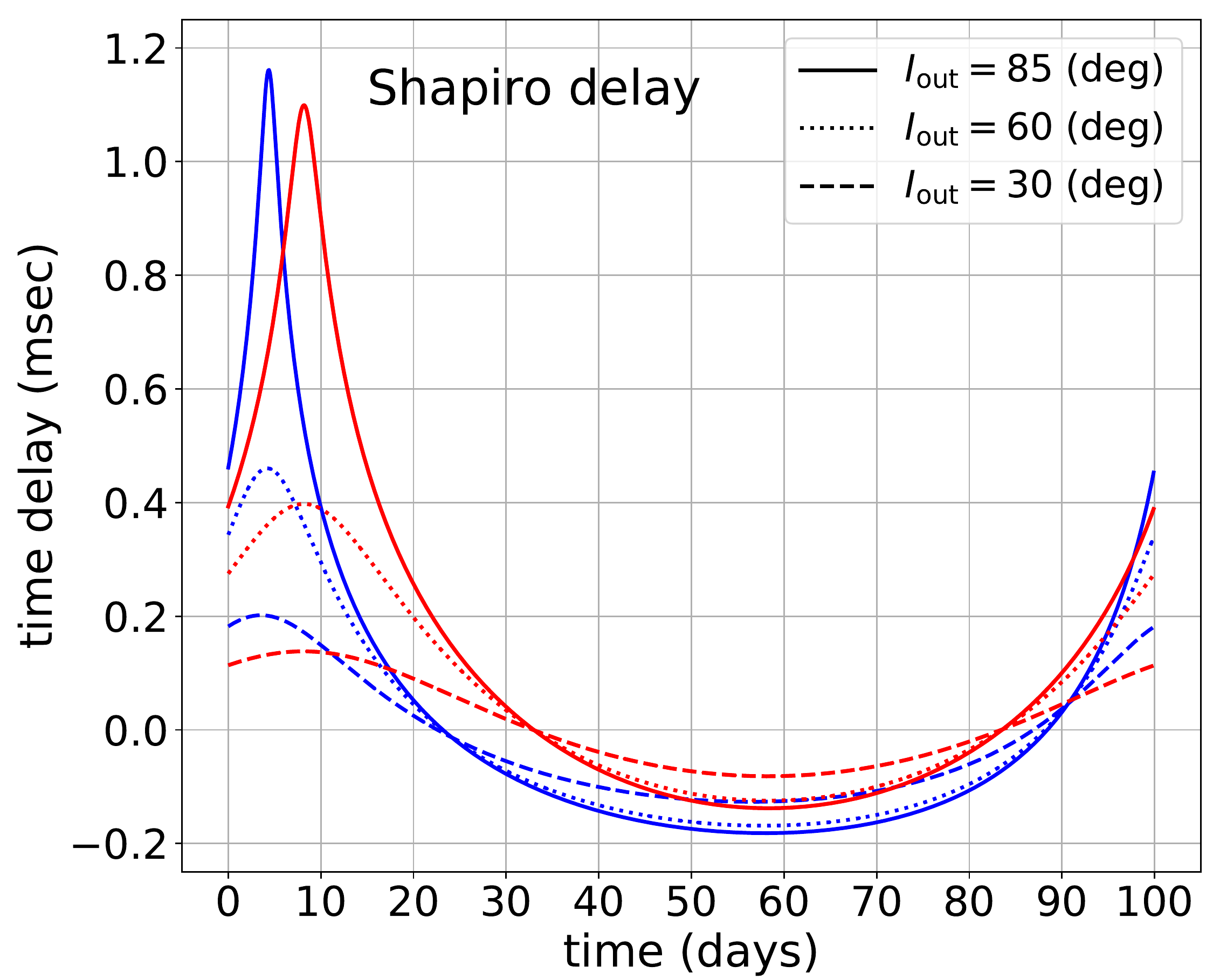}
\hspace{15pt}
\includegraphics[clip,width=8.0cm]{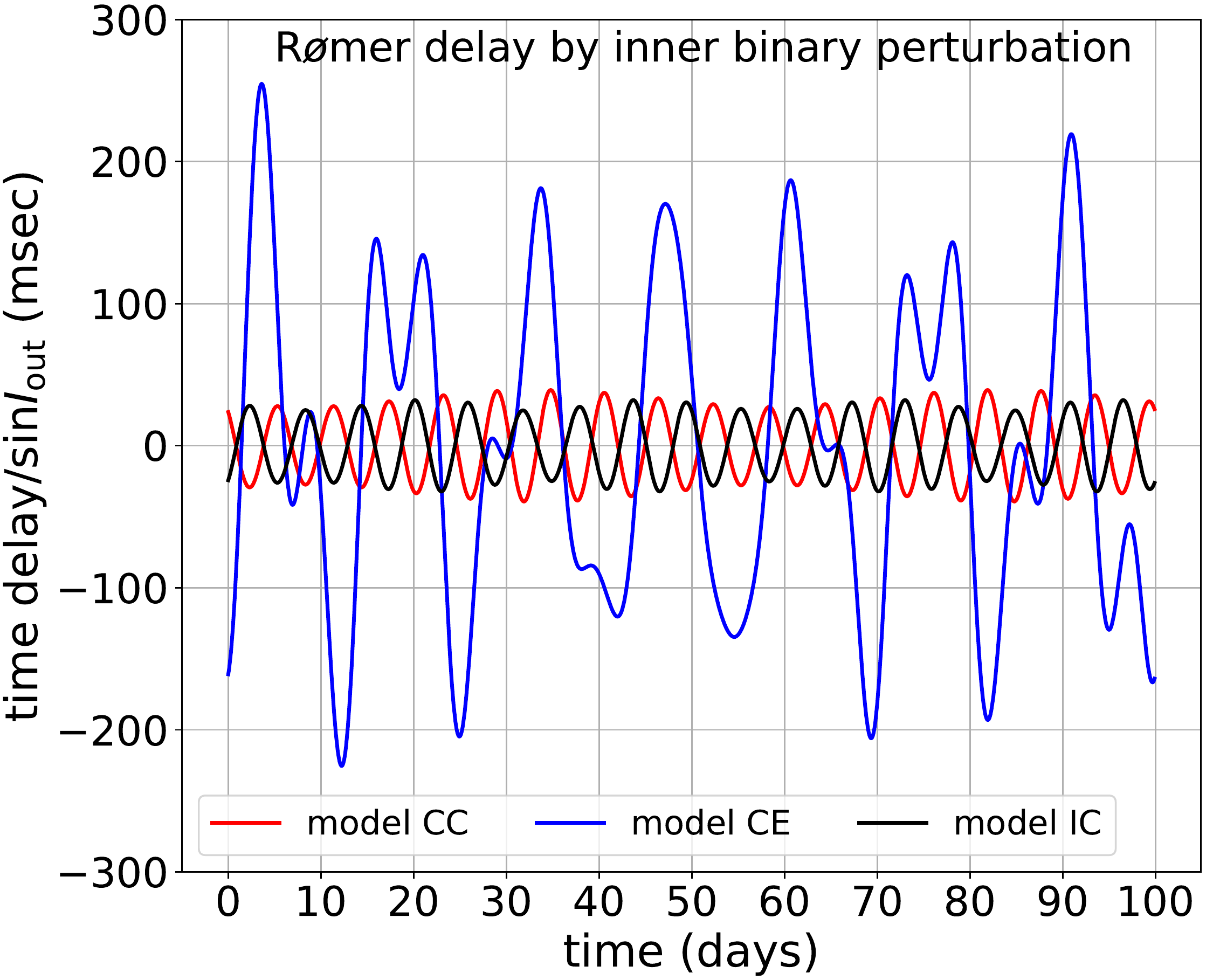}
\end{center}
\caption{The time-delay curves of the R\o mer delay by the outer Keplerian motion (upper-left), the Einstein delay (upper-right), the Shapiro delay (lower-left), and the R\o mer delay induced by the inner binary perturbation (lower-right) for the three models listed in Table \ref{tab:models}, respectively. The R\o mer delay by the inner-binary perturbation only shows the time-delay modes associated with the inner-binary frequency $\nu_\ii$. \label{fig:time-delay}}
\end{figure*}

Figure \ref{fig:time-delay} plots the time-delay curves for the
  three models. The upper-left, upper-right, and lower-left panels
  show the R\o mer, Einstein, and Shapiro delays, respectively, due to
  the outer Keplerian motion of a tertiary pulsar of period $P_\oo$.
  Strictly speaking, the outer orbit is perturbed by the inner-binary
  motion as well, but we neglect such small perturbations in those
  three panels for simplicity. Therefore these three time-delays are
  computed from equations (\ref{eq:Roemer}), (\ref{eq:Einstein}), and
  (\ref{eq:Shapiro}).  The perturbed outer Keplerian motion
  has been extensively discussed in Paper II, and the result (with
  some improvement relative to Paper II) is given in
  Appendix.

Thus the unseen inner-binary parameters are encoded only in the
  R\o mer delay modulation plotted in the lower-right panel of Figure
  \ref{fig:time-delay}, which is computed from equation
  (\ref{eq:Roemer-BBH}) for model CC, and the analytic perturbative
  formulae derived in \citet{Morais2011} for models CE and IC.

The upper-left panel of Figure \ref{fig:time-delay} indicates
  that the outer eccentricity $e_\oo$ distorts the sinusoidal curve of
  the R\o mer delay to some extent, according to equation (\ref{eq:Roemer}).
  Note that the constant offset in the figure is not relevant, and the eccentricity changes the shape of the delay. On the other hand, the Einstein delay
  (upper-right panel) is very sensitive to $e_\oo$ since equation
  (\ref{eq:Einstein}) is proportional to $e_\oo$.

Since the Shapiro delay is especially sensitive to the
  inclination of the outer orbit relative to the observer, the
  lower-left panel plots three different cases for $I_\oo$; nearly
  edge-on ($85$ deg), moderately inclined ($60$ deg), and nearly
  face-on ($30$ deg) for each model. While the amplitude of the
  Shapiro delay is smaller than the R\o mer delay and the Einstein
  delay for an eccentric orbit, it provides a unique constraint on
  $I_\oo$, especially for an inclined outer orbit, that is useful to
  break the parameter degeneracy as emphasized in the previous
  section.

\begin{deluxetable}{lcccccccccc}
\tablecolumns{4}
\tablewidth{1.0\columnwidth} 
\tablecaption{Models of non-circular/non-coplanar triples} 
\tablehead{ &  $e_\ii$ & $e_\oo$ & $i_\mathrm{mut}$ (deg)}
\startdata
CC (coplanar circular)  & 0.0 & 0.01 & 0.0 \\
CE (coplanar eccentric)  & 0.2 & 0.3 & 0.0 \\
IC (inclined circular) & 0.0 & 0.01 & 45 \\
\hline
\hline
\enddata
\label{tab:models}
\tablecomments{We adopt the same values for the other triple
  parameters:
  $m_1=$ $m_2=10~M_\odot$, $m_3=1.4~M_\odot$. $P_\ii=10$ days,
  $P_\oo=100$ days.  The angles are $\omega_\ii=30$ deg,
  $\omega_\oo=60$ deg, $\Omega_\ii=\Omega_\oo=0$ deg, and the
  initial true anomalies $f_\ii=120$ deg and $f_\oo=0$ deg.}
\end{deluxetable}

Finally the lower-right panel, the R\o mer delay due to the
  inner-binary perturbation, exhibits a clear shorter-term modulation
  (of period $\approx P_\ii/2$) periodicity, whose shape is also
  sensitive to the inner eccentricity and inclination.  Therefore, the
  detection of such short-term periodic features in the time delay
  component is a promising probe of the possible inner BBH of the
  unseen companion of the pulsar.

\section{Application to the existing binary neutron stars
    as a proof of concept to constrain an unseen inner binary
  \label{sec:application}}

Our methodology proposed in the present paper requires target
  pulsar binary systems with an unseen massive companion. The
  detection of the R\o mer delay modulation much shorter than the
  pulsar's Keplerian orbital period is an unambiguous proof that the
  unseen companion is indeed a binary, instead of a single
  object. While no interesting candidate is known for which our
  methodology can be applied, we attempt to put constraints on a
  possible inner binary for a companion of existing double neutron
  star (DNS) binaries through available pulsar arrival timing data.

\subsection{Pulsar arrival timing constraints}

Table \ref{tab:DNS} summarizes the current list of known DNS systems
with their orbital parameters derived from the pulsar timing
observations. Since they are interpreted as a binary system, $m_{12}$
and $m_{123}$ in Table \ref{tab:DNS} correspond to the companion mass
of the pulsar, and the total mass of the system, respectively. Since
both $m_{12}$ and $m_{123}-m_{12}$ are in the typical mass range of
neutron stars, it is very unlikely that those companions are binaries
of white dwarfs or black holes.  Nevertheless, the null detection of
the R\o mer delay modulation due to the inner binary motion within the
root mean square of the residuals $\sigma_\mathrm{rms}$ (the eighth
column of Table \ref{tab:DNS}) can put observational constraints on
the inner binary.

In the practical analysis of the pulsar timing, however, the
  signals from a triple may be degenerated with other parameters on a
  pulsar, which may obscure the interpretation or even the presence of
  the inner BBH if one relies on the standard pipeline that neglects
  the possible triple effects.  Therefore, an improved analysis taking
  account of such effects is required to constrain the system in a
  more quantitative and reliable fashion. This is beyond the scope of
  the present paper, but we are working on this problem and plan to
  show the result elsewhere (Kumamoto, Hayashi, Takahashi, and Suto,
  in preparation).

Because this is intended to be a merely proof-of-concept analysis, we
simply constrain those systems by assuming a coplanar near-circular
inner binary. Then using equation (\ref{eq:Roemer-BBH}), the inner
orbital period $P_\ii$ is constrained as
\begin{eqnarray}
\frac{K_\BBH P_\ii}{4\pi c} \sin{I_\oo} &=& 
\frac{1}{2}\frac{m_1m_2}{m_{12}^2} \left(\frac{m_{12}}{m_{123}}\right)^{2/3}
\left(\frac{P_\ii}{P_\oo}\right)^{7/3}x < \sigma_\mathrm{rms}.
\end{eqnarray}
If we further assume an equal-mass inner binary ($m_1=m_2$), the above
inequality reduces to
\begin{eqnarray}
\label{eq:const}
\frac{P_\ii}{P_\oo} < \left(\frac{8\sigma_\mathrm{rms}}{x}\right)^{3/7}
\left(\frac{m_{123}}{m_{12}}\right)^{2/7} \approx 0.009
\left(\frac{\sigma_\mathrm{rms}}{10~\mu \mathrm{sec}}\right)^{3/7}
\left(\frac{x}{5~\mathrm{sec}}\right)^{-3/7}
\left(\frac{m_{123}}{m_{12}}\right)^{2/7}.
\end{eqnarray}
The second column of Table \ref{tab:const} corresponds to the
  above upper limit on $P_\ii$ for a hypothetical equal-mass inner
  binary in a coplanar near-circular triple.  Those upper limits on
  $P_\ii$ are typically less than an hour, implying the future pulsar
  timing observation for pulsar -- massive BH binary candidates, if
  discovered in future, would strongly constrain the unseen inner
  binary in a similar fashion.

\begin{deluxetable}{lcccccccccc}
\tablecolumns{11}
\tablewidth{1.0\columnwidth} 
\tablecaption{List of known double neutron star systems} 
\tablehead{ System & $P_\mathrm{pulse}$ (msec) & $P_\oo$ (days) &
  $e_\oo$ & $x$ (sec) & $m_{123}~(M_\odot)$ & $m_{12}~(M_\odot)$ &
  $\sigma_\mathrm{rms}~(\mathrm{\mu sec})$ & $T_\mathrm{obs}$ (yrs) &
  Ref.}
\tabletypesize{\footnotesize}
\startdata
J0453+1559 & 45.8 & 4.072 & 0.113 & 14.467 & 2.733 & 1.174 & 3.88 & 2.4 & (1)\\ 
J0509+3801 & 76.5 & 0.380 & 0.586 & 2.051 & 2.80 & 1.46 & 102.36 & 3.1 & (2) \\
J0737-3039A & 22.7 & 0.102 & 0.088 & 1.415 & 2.587 & 1.249 & 54 & 2.7 & (3)\\
J0737-3039B & 2773 & 0.102 & 0.088 & 1.516 & 2.587 & 1.249 & 2169 & 2.7 & (3)\\
J1411+2551 & 62.5 & 2.616 & 0.170 & 9.205 & 2.538 & $>0.92$ & 32.77 & 2.4 & (4)\\
J1518+4904 & 40.9 & 8.634 & 0.249 & 20.044 & 2.718 & $>1.55$ & 6.05 & 13 & (5)\\
B1534+12 & 37.9 & 0.421 & 0.274 & 3.729 & 2.679 & 1.346 & 4.57 & 22 & (6)\\
J1753-2240 & 95.1 & 13.638 & 0.304 & 18.115 & -- & $>0.4875$ & 400 & 1.8 & (7)\\
J1756-2251 & 28.5 & 0.320 & 0.181 & 2.756 & 2.571 & 1.230 & 19.3 & 9.6 & (8)\\
J1757-1854 & 21.5 & 0.184 & 0.606 & 2.238 & 2.733 & 1.3946 & 36 & 1.6 & (9)\\
J1811-1736 & 104.2 & 18.779 & 0.828 & 34.783 & 2.57 & $>0.93$ & 851.2 & 7.6 & (10)\\
J1829+2456 & 41.01 & 1.176 & 0.139 & 7.237 & 2.606 & 1.310 & 10.086 & 17 & (11)\\
J1913+1102 & 27.3 & 0.206 & 0.090 & 1.755 & 2.89 & 1.27 & 56 & 7.3 & (12)\\
B1913+16 & 59.0 & 0.323 & 0.617 & 2.342 & 2.828 & 1.3867 & -- & 29 & (13)\\
J1930-1852 & 185.5 & 45.060 & 0.399 & 86.890 & 2.59 & $>1.30$ & 29 & 2.1 & (14)\\
J1946+2052 & 17.0 & 0.078 & 0.064 & 1.154 & 2.50 & $>1.18$ & 95.04 & 0.2 & (15)\\
\hline
\hline
\enddata
\label{tab:DNS}
\tablecomments{The values are based on: (1) \citet{Martinez2015} (2)
  \citet{Lynch2018} (3) \citet{Kramer2006} (4) \citet{Martinez2017}
  (5) \citet{Janssen2008} (6) \citet{Fonseca2014} (7)
  \citet{Keith2009} (8) \citet{Ferdman2014} (9) \citet{Cameron2018}
  (10) \citet{Corongiu2007} (11) \citet{Haniewicz2020} (12)
  \citet{Ferdman2020} (13) \citet{Weisberg2005} and \citet{Taylor1982}
  (14) \citet{Swiggum2015} (15) \citet{Stovall2018}}
\end{deluxetable}

\begin{deluxetable}{lcccc}
\tablecolumns{5}
\tablewidth{1.0\columnwidth} 
\tablecaption{Constraints on DNS Systems} 
\tablehead{ System & $P_{\ii,\mathrm{max}}$ (hrs)
  &$h(4\pi/P_{\ii,\mathrm{max}})$ & $t_\mathrm{merge}(P_{\ii,\mathrm{max}})$ (yrs) & $D$
  (kpc)}
\startdata
J0453+1559 & $0.463$ &$8.4\times10^{-22}$ & $3.9\times 10^6$ & 0.52 \\
J0509+3801 & $0.384$ & $1.0\times10^{-22}$ & $1.6\times 10^6$ & 7.08 \\
J0737-3039A & $0.0939$ & $1.2\times10^{-21}$ & $4.9\times 10^4$ & 1.17 \\
J1411+2551 & $0.946$ & $1.6\times10^{-22}$ & $3.9\times 10^7$ & 1.13 \\
J1518+4904 & $0.952$ & $4.5\times10^{-22}$ & $1.7\times 10^7$ & 0.96 \\
B1534+12 & $0.0878$ & $1.8\times10^{-21}$ & $3.6\times 10^4$ & 0.93 \\
J1753-2240 & -- & -- & -- & 6.93 \\
J1756-2251 & $0.143$ & $1.1\times10^{-21}$ & $1.6\times 10^5$ & 0.95 \\
J1757-1854 & $0.115$ & $7.5\times10^{-23}$ & $7.1\times 10^4$ & 19.6 \\
J1811-1736 & $15.5$ & $2.8\times10^{-24}$ & $6.6\times 10^{10}$ & 10.16 \\
J1829+2456 & $0.259$ & $8.5\times10^{-22}$ & $6.8\times 10^5$ & 0.91 \\
J1913+1102 & $0.180$ & $1.3\times10^{-22}$ & $2.7\times 10^5$ & 7.14 \\
B1913+16 & -- & -- & -- & 5.25 \\
J1930-1852 & $5.38$ & $4.1\times10^{-23}$ & $2.3\times 10^9$ & 2.48 \\
J1946+2052 & $0.100$ & $3.5\times10^{-22}$ & $6.5\times 10^4$ & 3.51 \\
\hline
\hline
\enddata
\label{tab:const}
\tablecomments{The upper limits of inner orbital period from the
  equation (\ref{eq:const}), and the corresponding GW strain $h$ (see
  equation (\ref{eq:strain})) and the merger time assuming equal-mass
  circular binaries. We assume minimum-mass companions for the systems
  that only the lower limits of companion masses are determined.  The
  fifth column denotes the mean values of the distances of the systems
  in \citet{Haniewicz2020}, which are mainly determined by the
  dispersion measures.}
\end{deluxetable}

\subsection{Low-frequency gravitational wave constraints}

If $P_\ii$ is sufficiently small, the inner binary is difficult
  to be distinguished from a single massive object from the tertiary
  motion.  On the other hand, the gravitational wave (GW) from such
  short-period compact binaries may become detectable. Thus the
  presence of the inner binary can be probed in a complementary
  fashion by combining the pulsar timing and the GW data. Indeed as we
  show below, if the existing DNS systems have an inner BBH whose
  orbital period is shorter than the pulsar timing constraint, their
  low-frequency GW may be detectable by future space-based GW
  missions.

For instance, LISA \citep[e.g.][]{LISA2017} whose launch is scheduled
in 2030's will be sensitive to very low-frequency GW signals down to
$\sim10^{-4}$ Hz.  Previous proposals and discussions to search for
the low-frequency GW sources with LISA include compact binary
\citep[e.g.][]{Nelemans2001}, ultrashort-period planet
\citep[e.g.][]{Cunha2018,Wong2019}, and the Kozai-Lidov oscillations
\citep[e.g.][]{Gupta2020}. We argue that an inner BBH in a triple that
cannot be ruled out by the pulsar timing will be significantly
constrained, or even detected by future space-based GW missions
including LISA, DECIGO \citep[e.g.][]{DECIGO2017}, and BBO
\citep[e.g.][]{BBO2006}.

In reality, the GW signals from an inner binary are also
  modulated in frequency and phase depending on the outer orbit
  parameters. Therefore, the precise detection of the inner BBHs in
  triple systems requires to take account of such triple effects
  simultaneously. In the following calculation, however, we simply
  assume that the outer orbital parameters are determined with high
  precision and properly subtracted from the entire signals. This is yet another reason why the following results should be regarded as a
  proof-of-concept example. Nevertheless, they present an interesting
  possibility to search for possible BBHs embedded in triple systems.
  Indeed the GW analysis is independent of the nature of the tertiary
  body except for its gravity. Thus the future improved GW analysis
  including the triple effects may detect star-BBH, pulsar-BBH, and
  BH-BBH triple systems as well, if they exist at all.

The characteristic amplitude of the GW strain emitted from
a circular binary is \citep[e.g.][]{Hartle2003,Schutz2009}
\begin{eqnarray}
\label{eq:strain}
h(\nu) &\sim& \frac{1}{a_\ii(\nu) D}\left(\frac{2\mathcal{G}m_1}{c^2}\right)
\left(\frac{2\mathcal{G}m_2}{c^2}\right)
= \frac{2^{4/3}\mathcal{G}^{5/3}\mathcal{M}^{5/3}}{Dc^4} \nu^{2/3}
\\
&\approx& 7.7\times 10^{-20} \left(\frac{\mathcal{M}}{8.7~M_\odot}\right)^{5/3}
\left(\frac{P_\ii}{0.01~\mathrm{days}}\right)^{-2/3}
\left(\frac{D}{1~\mathrm{kpc}}\right)^{-1},
\end{eqnarray}
where $D$ is the distance to the system, $\nu$ is the GW frequency, which corresponds to $4\pi/P_\ii$ for a circular binary, and $\mathcal{M}$ is the
chirp mass of the binary:
\begin{eqnarray}
  \mathcal{M} \equiv \frac{(m_1m_2)^{3/5}}{m_{12}^{1/5}}\approx 8.7~M_\odot
  \left(\frac{m_1m_2}{100~M^2_\odot}\right)^{3/5}
  \left(\frac{m_{12}}{20~M_\odot}\right)^{-1/5}.
\end{eqnarray}

\begin{figure*}
\begin{center}
\includegraphics[clip,width=8.5cm]{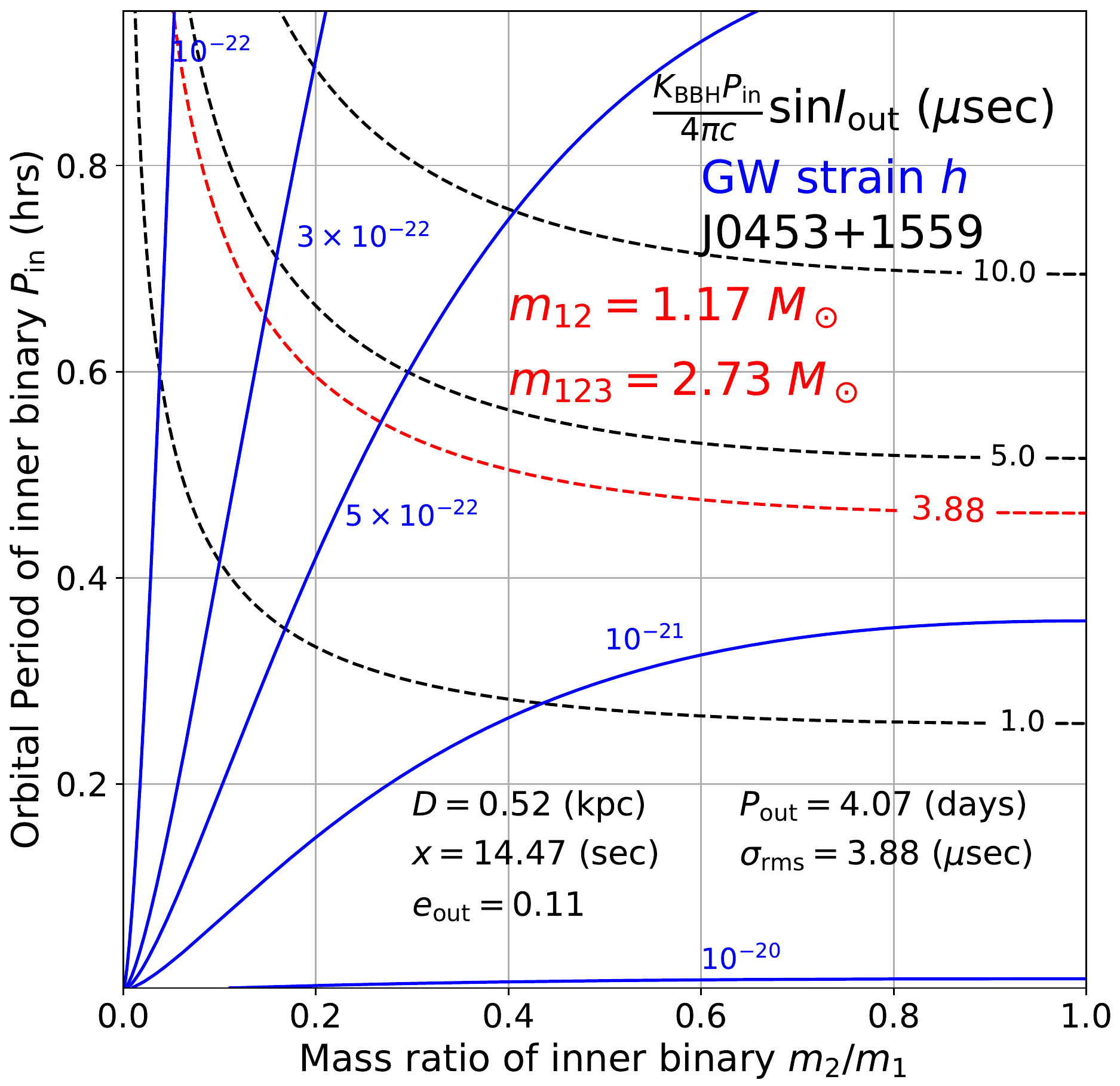}
\hspace{15pt}
\includegraphics[clip,width=8.5cm]{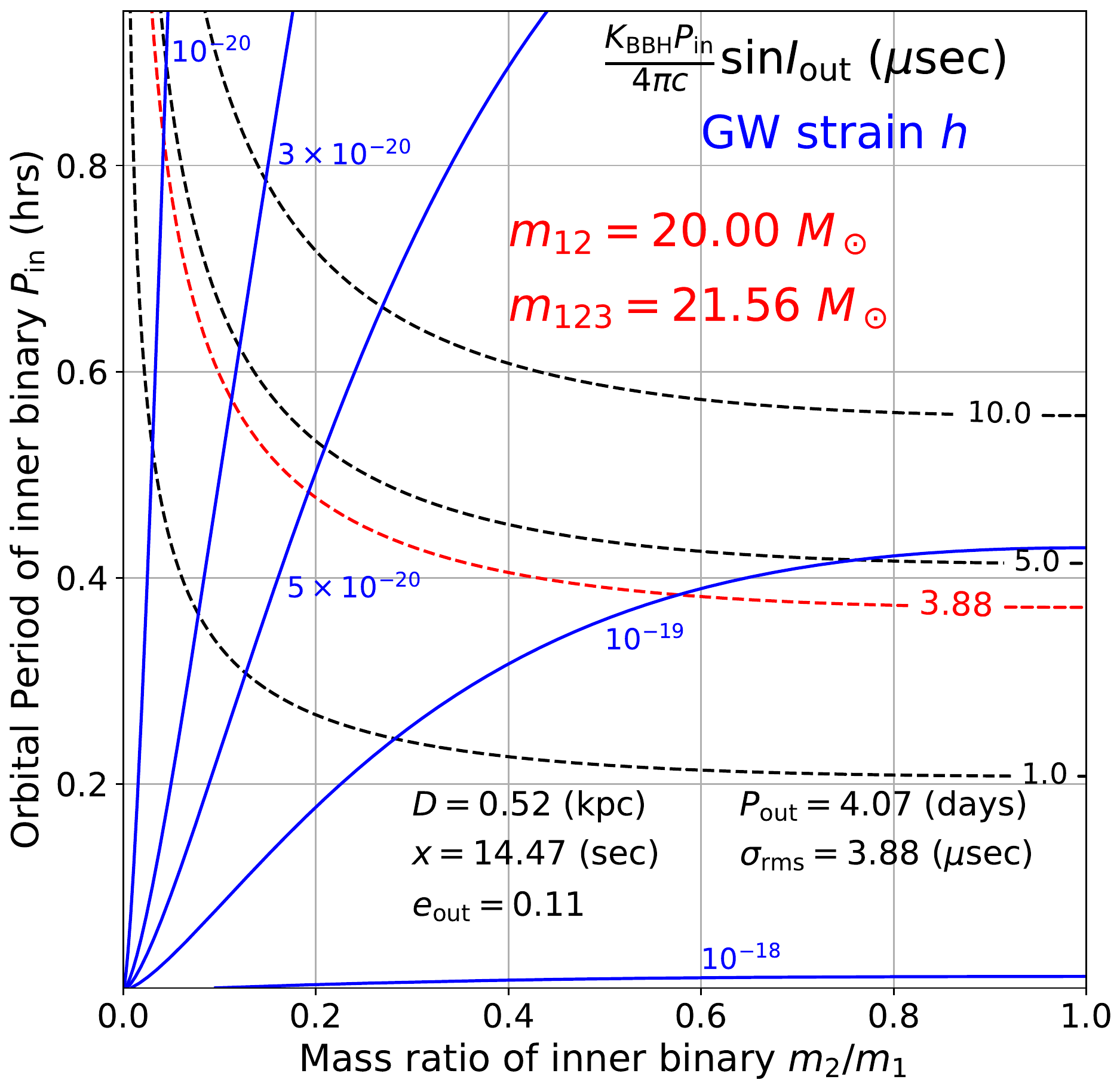}
\end{center}
\caption{Characteristic amplitudes of the R\o mer delay induced by the
  inner-binary perturbation (dashed black and red lines) and the
  gravitational wave strain $h$ (solid blue line) as the function of
  inner-binary orbital period $P_\ii$ and mass ratio $m_2/m_1$.  The left and right
  panels show the constraints for J0453+1559 and a hypothetical inner
  BBH with $m_{12}=20~M_\odot$, respectively. \label{fig:constraint}}
\end{figure*} 

As a specific example, we consider the DNS binary J0453+1559, and
  assume that it is indeed a triple with the unseen companion being a
  coplanar near-circular inner compact binary of the mass ratio
  $m_2/m_1$ and the orbital period $P_\ii$, instead of a single
  neutron star.  The left panel of Figure \ref{fig:constraint} plots
  the amplitudes of the corresponding GW strain (thick solid lines)
  and the R\o mer delay modulation (thin dashed lines). The region
  above the red line $P_\ii >P_{\ii,\mathrm{max}}$ is excluded because
  it should exhibit the arrival time modulation larger than the
  observed $\sigma_{\rm rms}$. Interestingly, the region below the
  limit indicates that the GW strain at the frequency corresponding to
  $4\pi/P_{\ii,\mathrm{max}}$ amounts to $h>10^{-21}$ that may be
  detectable including LISA, DECIGO, and BBO; see Figure \ref{fig:sensitivity} below.

Assuming that each DNS system has an equal-mass circular inner binary, we can put rough constraints on the properties of the possible inner binaries. The GW emission merger time $t_\mathrm{merge}(P_\ii)$ for an equal-mass circular binary is \citep[][]{Peters1964}
\begin{eqnarray}
t_\mathrm{merge}(P_\ii) = \frac{5}{64} \left(\frac{\mathcal{G} m_{12}}{c^3}\right)^{-5/3} \left(\frac{P_\ii}{2\pi}\right)^{8/3}
\approx 3.92 \times 10^{7} \left(\frac{P_\ii}{\mathrm{hrs}}\right)^{8/3}
\left(\frac{m_{12}}{M_\odot}\right)^{-5/3}~\mathrm{yrs}.
\end{eqnarray}
Table \ref{tab:const} summarizes the upper limit on the inner binary period $P_{\ii,\mathrm{max}}$, the corresponding GW strain $h(4\pi/P_{\ii,\mathrm{max}})$, and the GW emission merger time $t_\mathrm{merge}(P_{\ii,\mathrm{max}})$.

In order to examine the feasibility to constrain the inner BBH in
  triple systems, we assume exactly the same triple parameters for
  the DNS binary J0453+1559 except that the inner companion mass is
  $m_{12}=20~M_\odot$. The amplitudes of the R\o mer delay modulation
  and the GW strain for the hypothetical system are plotted in the
  right panel of Figure \ref{fig:constraint}.  Since the GW strain
  becomes about two orders of magnitude larger compared with the left
  panel, an inner BBH, if exists at all, would be detected for almost
  all the the parameter space with either the precise high-cadence
  pulsar timing or the low-frequency GW observation.

Figure \ref{fig:sensitivity} plots the characteristic GW strain
  $h(4\pi/P_{\ii})$ for hypothetical circular inner binaries; long-dashed,
  short-dashed, dotted and dash-dotted lines corresponds to the inner
  BBHs of $(m_1, m_2)$ $=(30~M_\odot, 30~M_\odot)$, $(10~M_\odot,
  10~M_\odot)$, $(15~M_\odot, 5~M_\odot)$, and $(5~M_\odot, 5~M_\odot)$,
  located at $D=10$ kpc from us. Solid lines indicate $h(4\pi/P_{\ii})$ from
  the existing DNS systems shown in Table \ref{tab:const}, assuming
  an equal-mass circular inner binary instead of the companion neutron
  star. We also show the expected LISA sensitivity curve in 4 year
  mission \citep{Robson2019}, the expected sky-averaged sensitivity curves of DECIGO and BBO \citep[][]{Yagi2011,Yagi2017}, and the aLIGO sensitivity curve from a technical note T1800044-v5 \citep[][]{Barsotti2018}. The figure shows that very short-period inner binary companions are already excluded by non-detection with aLIGO, and the other space-based missions (LISA, DECIGO, and BBO) have enough sensitivity to detect an inner binary with hr-scale orbital period in the future.

Clearly, the joint analysis of the pulsar timing and GW
  observation is very effective, and can constrain the presence of an
  unseen inner binary in a complementary fashion; inner BBHs with a
  shorter orbital period that cannot be probed by the pulsar timing
  analysis will be detectable by future space-based GW detectors such as LISA, DECIGO, and BBO. Additionally, very short-period inner binaries with sub-second orbital periods can be already probed or constrained by current ground-based GW detectors such as aLIGO.  

\begin{sidewaysfigure*}
\begin{center}
\includegraphics[clip,width=23.0cm]{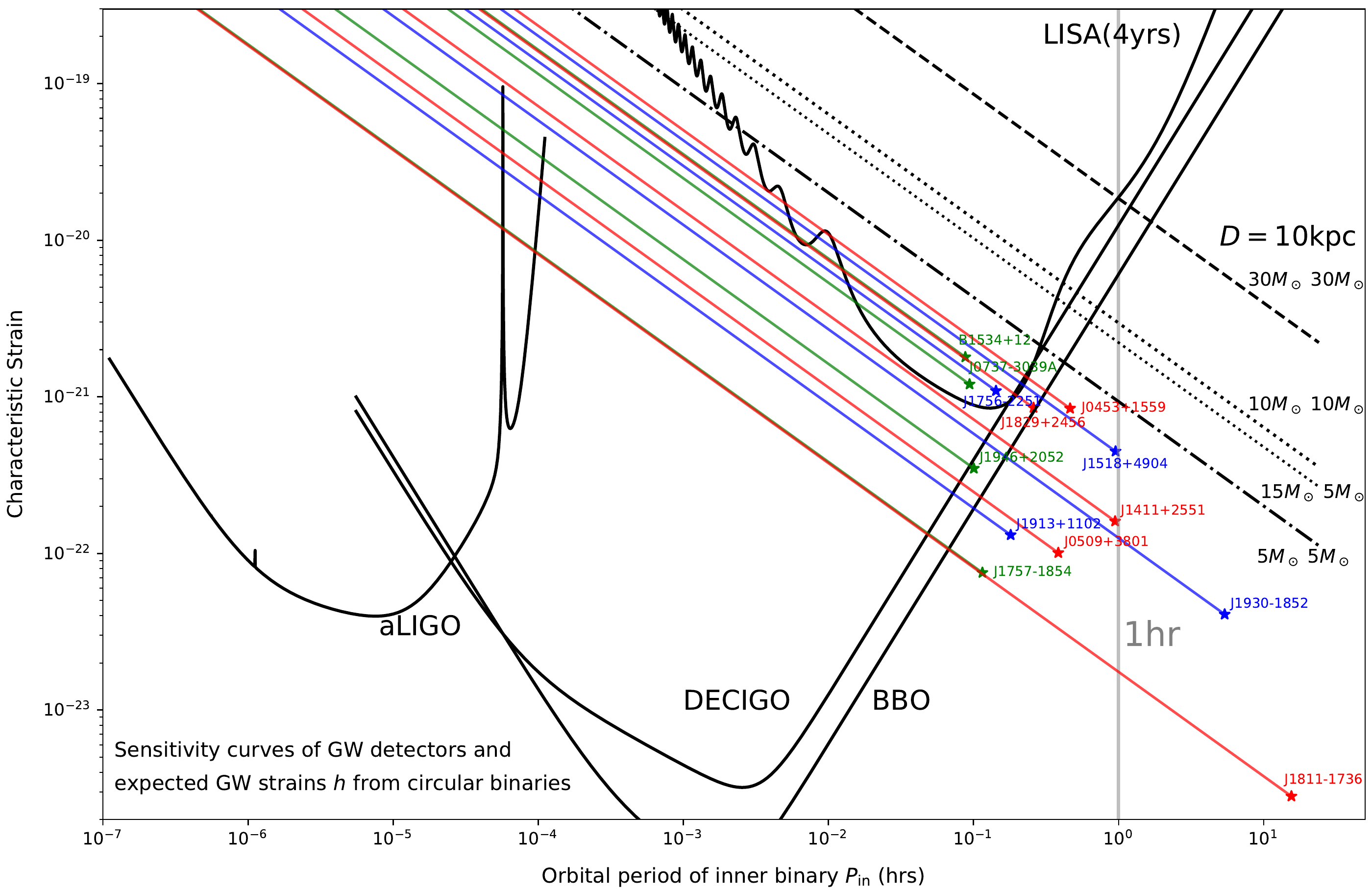}
\end{center}
\caption{The sensitivity curves of three future space-based GW
  detectors (LISA, DECIGO, and BBO) and a present ground-based GW
  detector (aLIGO). The characteristic GW strains $h(4\pi/P_\ii)$ are
  also plotted for hypothetical inner circular BBHs located at $10$
  kpc, and assumed equal-mass circular inner binary companions in the
  currently known DNS systems. The stars represent the upper limit of
  inner orbital periods and corresponding GW strains (see Table
  \ref{tab:const}). The DNS systems J1753-2240 and B1913+16 are not
  plotted in this figure since no constraints are available for these
  systems. All the results presented here assume that the GW
    modulations in frequency and phase by the outer orbit are properly
    subtracted by the predetermined outer orbital parameters.}
\label{fig:sensitivity} 
\end{sidewaysfigure*} 


\clearpage
\section{Conclusion \label{sec:conclusion}}
According to the proposed formation scenarios of binary black holes (BBHs), there should be abundant wide-orbit progenitor BBHs in the universe. In the previous papers (Papers I and II), we examine the feasibility to identify such BBHs in triples by detecting short- and long-term radial velocity modulations of a tertiary star induced by the inner-binary perturbation.

In this paper, we consider a pulsar-inner BBH triple, and propose a methodology to search for an inner hidden BBH on the basis of the pulsar arrival time analysis.
While the presence of such triples is currently uncertain, if such a triple exists, high precision and cadence pulsar timing can clearly detect a BBH inside distant triples beyond several kpc, inaccessible with the radial velocity observation.

We show that an inner hidden BBH can be identified by the short-term
R\o mer delay modulation by the inner-binary perturbation with roughly half an inner-binary orbital period. The analytic expression reveals that the modulation has sufficient amplitude for the detection up to $\sim 50$ msec, for our fiducial case of an equal-mass coplanar near-circular inner BBH with $m_{12}=20~M_\odot$, $P_\ii=10$ days, and $P_\oo=100$ days. 
We also show that the mass of each body and orbital parameters of a triple can be unambiguously determined combining the relativistic time delays, especially the Shapiro delay up to $\sim$ msec for a nearly edge-on system.

As an application of our methodology, we perform a proof-of-concept analysis, and put rough constraints on possible inner binary companions of existing double neutron-star binaries (DNS) using the root mean square of the residuals in observational arrival timing data. 
We find that our proposed methodology has the sensitivity even on inner binaries with relatively short orbital periods down to $\sim$ hrs when $P_\oo < O(1)$ day and the pulsar timing precision is on the order of $\mu$sec. While it is unlikely that existing DNS systems actually have inner binary companions instead of singles, the result indicates that our methodology can probe various inner binaries, which may be hardly detected with radial velocity observation.

Additionally, the result implies the possibility of our methodology as
a complementary method to the direct gravitational wave (GW)
observations.  Inner BBHs with $\lesssim 1$ hr orbital period located
at $\lesssim 10$ kpc produce the GW strains detectable by future
space-based GW detectors including LISA, DECIGO, and BBO, and very
short-period inner BBHs with sub-second orbital periods can be already
probed by the current ground-based GW detectors such as aLIGO.
Therefore, the combination of our methodology with current and
  future GW observations will provide a potential technique to search
  for inner BBHs in the near future.

Although we concentrate on the short-term modulation induced by the
inner-binary perturbation in this paper, the long-term effects on the time delay such as the Kozai-Lidov oscillations in
non-coplanar triples would also provide the feasible methodology to detect inner BBHs, as same as the radial velocity (see Paper II).

The dynamics of triple systems we applied in this paper is quite
generic, and can be applied to various other interesting systems,
apart from a pulsar-inner BBH triple. For instance, the formation of
binary planets in exoplanetary systems via gravitational scattering is
predicted theoretically \citep[e.g.][]{Ochiai2014}, and the search
through the transit observation is currently proposed
\citep[e.g.][]{Lewis2015}.  Although the modulation of the tertiary
motion induced by a binary planet should be generally small, high
precision pulsar timing may be possible to detect the tiny modulation
if a tertiary is a pulsar. Adapting our methodology to searching for a
binary planet around a pulsar will be one possible application.

In the near future, the space-based GW detectors will start
  operating and open a new window for the compact object survey.  In
  Figure \ref{fig:sensitivity}, we show a possible constraint on
  assumed inner binaries inside triples from the upcoming such
  detectors. The non-trivial assumption underlying the plot is that
  the GW modulation due to the outer orbit can be precisely
  subtracted.  For that purpose, it is required to develop a data
  processing algorithm to detect the GW signals from triple systems.
  Obviously such a pipeline can be applied not only to pulsar-BBH
  triples that we have examined in the present paper, but also to BH
  triple systems.  For example, \citet{Gupta2020} considered the GW
  signals from an inner binary with a wide-separation tertiary
  undergoing the Kozai-Lidov oscillation. For more compact BH triple
  systems, however, the resulting GW signals would show more
  complicated behavior since the GWs from both inner and outer orbits
  may enter the detectors simultaneously, and it is challenging to
  disentangle the two components under the situation that three BHs
  are strongly interacted in tightly packed systems.  Nevertheless,
  such directions should be very important and rewarding since they
  would lead to the first detection of tight BH triples hidden somewhere in
  the universe.

Finally, we would like to emphasize that the methodology proposed here
is not just a theoretical idea, but an observationally feasible
methodology for searching for otherwise unseen astrophysical
objects. Theoretically, it is estimated that there are $10^8-10^9$
stellar mass black holes in our Galaxy \citep[e.g.][]{Shapiro1983},
and currently there are many proposals to search for star - black hole
binaries with Gaia\citep[e.g.][]{Kawanaka2016,Breivik2017,Mashian2017,
  Yamaguchi2018,Shikauchi2020_2} and TESS
\citep[e.g.][]{Masuda2019}. LIGO's continuous detection of merger
events \citep[e.g.][]{Abbott2016,Abbott2020} implies that abundant
black holes may form BBHs, and still stay in the universe before
coalescence. In the near future, the methodology proposed here becomes
practically useful in detecting such not-yet-known astronomical
objects.

\acknowledgments
We thank the referee, Scott Ransom, for careful reading of our
  manuscript and several valuable comments and suggestions.  We are
  highly grateful to Toshio Fukushima for important discussion on the
  dynamics of a triple system, and Mamoru Sekido for introducing us
  with basics of the pulsar timing analysis. We gratefully
acknowledge the support from Grants-in Aid for Scientific Research by
the Japan Society for Promotion of Science (JSPS) No.18H012 and
No.19H01947, and from JSPS Core-to-core Program ``International
Network of Planetary Sciences''.

\vspace{5mm}

\appendix
\section{The modulation on the Kepler motion induced
  by the inner-binary perturbation \label{sec:appendix}}

In order to clearly detect the R\o mer delay induced by the
inner-binary perturbation, it is important to precisely determine the
Keplerian motion and properly subtract it from the timing data. In
this section, we briefly discuss how the Keplerian motion is modulated
and the best-fit parameters of the Keplerian orbit can be interpreted
under the presence of the inner-binary perturbation. For simplicity,
we only consider a coplanar near-circular triple and extend the
discussion in our previous paper (Paper II).

Following \citet{Morais2008} and Paper II, the radial Keplerian
motion under the inner-binary perturbation is written by the unperturbed Keplerian term $z^{(0)}_\kep(t)$ and the tiny correction $\delta z_\kep(t)$ as
\begin{eqnarray}
\label{eq:V0Kep}
z_\kep(t) &=& z^{(0)}_\kep(t) + \delta z_\kep(t) \nonumber \\
&=& \frac{m_{12}}{m_{123}} a_\oo \left[1+\frac{3}{4}\frac{m_1m_2}{m_{12}^2}\left(\frac{a_\ii}{a_\oo}\right)^{2}
\right] \sin{I_\oo}\sin(\nu_\oo t+f_\ooz+\omega_\oo),
\end{eqnarray}
where $\nu_\oo$ and $\omega_\oo$ denote the mean motion and argument
of pericenter of the tertiary pulsar, and $f_\ooz$ is the initial true
anomaly at $t=0$. Since the outer orbit in a triple system should have
a non-vanishing eccentricity due to the inner-binary perturbation, $\omega_\oo$ in the above expressions is
generally well defined. The above expression shows the amplitude of
the radial Keplerian motion is modulated by the inner-binary
perturbation with the order of $(a_\ii/a_\oo)^2$.

The averaged orbital period $P^{(\mathrm{ave})}_\oo$ over a orbital
motion is also modulated due to the long-term effects of the
inner-binary perturbation. Since $\omega_\oo$ and $f_\ooz$ are not
constant with time under the perturbation, the angular frequency
corresponding to the averaged orbital period is written as
\begin{eqnarray}
\nu^{(\mathrm{ave})}_\oo \equiv \frac{2\pi}{P^{(\mathrm{ave})}_\oo} \approx \nu_\oo + \dot{\omega}_\oo + \dot{f}_\ooz.
\end{eqnarray}
The $\dot{\omega}_\oo$ and $\dot{f}_\ooz$ can be calculated by the
Lagrange planetary equations using the orbit-averaged quadrupole
Hamiltonian of triple system. The orbit-averaged quadrupole
Hamiltonian $\bar{F}$ is given by \citep[e.g.,][]{Morais2012}:
\begin{eqnarray}
\label{eq:F}
  \bar{F} &=& C_{\mathrm{quad}}[2-12e_\ii^2-6(1-e_\ii^2)
    (\sin{I_\ii}\sin{I_\oo}\cos(\Delta\Omega)+\cos{I_\ii}\cos{I_\oo})^2+30e_\ii^2
    \cr
    && \times(-\sin{I_\oo}\cos{I_\ii}\sin{\omega_\ii}\cos(\Delta\Omega)
    -\sin{I_\oo}\cos{\omega_\ii}\sin(\Delta\Omega)
    +\sin{I_\ii}\sin{\omega_\ii}\cos{I_\oo})^2],
\end{eqnarray}
where 
\begin{eqnarray}
\label{eq:C}
  C_{\mathrm{quad}} \equiv \frac{\mathcal{G}}{16}
  \frac{m_1m_2}{m_{12}}\frac{m_3}{(1-e_\oo^2)^{3/2}}
  \left(\frac{a_\ii^2}{a_\oo^3}\right) ~~~\mathrm{and}~~~
\Delta\Omega \equiv \Omega_\ii-\Omega_\oo.
\end{eqnarray}
The Lagrange planetary equations of $\omega_\oo$ and $f_\ooz$ are
\citep[see e.g.][]{Danby1988}
\begin{eqnarray}
\dot{\omega}_\oo &=&
  -\frac{\sqrt{1-e_\oo^2}}{\mu_\oo\nu_\oo a_\oo^2e_\oo}
  \frac{\partial \bar{F}}{\partial e_\oo}
  +\frac{\cos{I_\oo}}{\mu_\oo\nu_\oo a_\oo^2\sqrt{1-e_\oo^2}\sin{I_\oo}}
  \frac{\partial \bar{F}}{\partial I_\oo}
\end{eqnarray}
and
\begin{eqnarray}
\dot{f}_\ooz &\approx& \dot{\bar{M}}_\ooz =
  \frac{1-e_\oo^2}{\mu_\oo\nu_\oo a_\oo^2e_\oo}
  \frac{\partial \bar{F}}{\partial e_\oo}
  +\frac{2}{\mu_\oo\nu_\oo a_\oo}
  \frac{\partial \bar{F}}{\partial a_\oo},
\end{eqnarray}
where $\mu_\oo \equiv m_3m_{12}/m_{123}$.  The $\bar{M}_\ooz$ is
defined by the mean anomaly $M_\oo$ to avoid the secular term in the
equation as
\begin{eqnarray}
M_\oo \equiv \nu_\oo(t-t_0)+M_\ooz \equiv \int^t_{t_0}{\nu_\oo dt} + \bar{M}_\ooz,
\end{eqnarray}
where $t_0$ is the initial time.
Note that we use the fact that the initial true anomaly is well
approximated by the initial mean anomaly for a near-circular case.  Substituting the equation
(\ref{eq:F}) into the Lagrange planetary equations, and neglect the
outer eccentricity $e_\oo$, we obtain
\begin{eqnarray}
  \dot{\omega}_\oo \approx \frac{12C_\mathrm{quad}}{\mu_\oo \nu_\oo a_\oo^2}
  ~~~\mathrm{and}~~~\dot{f}_\ooz
  \approx \frac{12C_\mathrm{quad}}{\mu_\oo \nu_\oo a_\oo^2}.
\end{eqnarray}
Therefore, the averaged orbital period $P^{(\mathrm{ave})}_\oo$ over a
orbital motion is approximated as
\begin{eqnarray}
\label{eq:Pave}
\frac{P^{(\mathrm{ave})}_\oo}{P_\oo}
\approx \frac{2\pi}{\nu_\oo+\dot{\omega}_\oo+\dot{f}_\ooz}
\approx 1-\frac{2\dot{\omega}_\oo P_\oo}{2\pi}.
\end{eqnarray}
Incidentally, Figure 4 in Paper II indicated the factor 2 difference between $P_\oo$ and $P^{(\mathrm{ave})}_\oo$, but it is due to the additional contribution from $\dot{f}_\ooz$ that was omitted in Paper II. Thus the above equation (\ref{eq:Pave}) correctly reproduces the numerical result from analytic computation.


\end{document}